\documentclass[alpha-refs]{wiley-article}

\usepackage{color}
\usepackage{ulem}

\usepackage{siunitx}

\papertype{Original Article}
\paperfield{Journal Section}

\title{Evaluating Anisotropy-based Monin-Obukhov Similarity Theory over Canopies and Complex Terrain}

\author[1]{Tyler S. Waterman}
\author[2]{Ivana Stiperski}
\author[3]{Nathaniel Chaney}
\author[1]{Marc Calaf}

\affil[1]{Mechanical Engineering, University of Utah, Salt Lake City, Utah, United States of America}
\affil[2]{Atmospheric and Cryospheric Sciences, Universität Innsbruck, Innsbruck, Tyrol, Austria}
\affil[3]{Civil and Environmental Engineering, Duke University, Durham, North Carolina, United States of America}

\corraddress{Tyler Waterman PhD, Mechanical Engineering, University of Utah, Salt Lake City, Utah, United States of America}
\corremail{tyswater@gmail.com}

\fundinginfo{National Science Foundation (NSF), Atmospheric and Geospace Science (AGS), Award Number: 2412560, 2414424; European Research Council (ERC) under the European Union’s Horizon 2020 research and innovation program, Grant/Award Number: 101001691}

\runningauthor{Waterman}
\usepackage[displaymath, mathlines]{lineno}

\begin{document}

\maketitle

\begin{abstract}

\end{abstract}

\section*{Data Availability}
The data that support the findings of this study are openly available in National Ecological Observation Network Data Portal at https://data.neonscience.org/data-products/DP4.00200.001/, reference number DOI:10.48443/r7zp-y487.

\section*{Abstract}
Monin Obukhov Similarity Theory (MOST) has long served as the basis for parameterizations of turbulence exchange between the surface and the atmospheric boundary layer in models for weather and climate prediction. Decades of research, however, has illuminated some of the limitations of MOST based surface layer parameterizations, particularly when MOST’s foundational assumptions of flat and horizontally homogeneous terrain are violated. Recent work has leveraged the anisotropy of turbulence as an additional non-dimensional term to extend and generalize MOST to complex terrain.
\par In this work, we examine the performance of this generalized MOST for the scaling of velocity variances, refit these scalings, and study key characteristics of turbulence anisotropy across the 47 towers in the wide ranging National Ecological Observation Network (NEON). NEON in particular covers a diverse selection of ecosystems, from the arctic circle to tropical islands, and as such expands the previous generalized MOST to vegetated canopies and other environments not examined in previous studies. The work finds that anisotropy generalized MOST readily extends to these new environments, with robust performance across seasons over a wide range of canopy and terrain configurations. Results also further illuminate velocity variance scaling across degrees and forms (one-component and two-component) of anisotropy, relate the degree of anisotropy to key environmental characteristics, and evaluate scaling differences between canopies and non-vegetated surfaces across atmospheric conditions, seasons and the diurnal cycle. Overall, this study expands our understanding of the complexity of turbulence while paving the way for improved surface layer parameterizations. 
\par \textbf{Keywords: } similarity scaling, velocity variances, surface layer scaling, boundary-layer meteorology, turbulence anisotropy

\section{Introduction}
Atmospheric surface layer (ASL) exchange between the land surface, atmospheric boundary layer (ABL) and broader atmosphere plays a fundamental role in the regulation of local and global water, energy and carbon cycles. The coupling of a complex, heterogeneous surface with the overlying turbulent atmosphere is fundamental for understanding, analyzing, modeling, and predicting weather and climate. Our understanding of ASL turbulence remains incomplete outside of very idealized conditions, which in turn affects the performance of our models of ABL turbulence, exchange between the surface and atmosphere, and atmospheric phenomena more broadly \citep{calaf_boundary-layer_2023,bou-zeid_persistent_2020,mauder_surface-energy-balance_2020,cohen_review_2015,otarola_bustos_subgrid_2023,srivastava_note_2021,banks_sensitivity_2016,akylas_sensitivity_2007}. The theories often applied for ASL parameterization schemes assume stationary and horizontally homogeneous flows, however these assumptions are frequently violated under real world conditions. This includes Monin Obukhov Similarity Theory (MOST) \citep{monin_basic_nodate}, the primary scaling theory applied in both ABL literature and in ABL schemes in atmospheric models, such as numerical weather prediction (NWP), and Earth system models (ESMs). The simplifying assumptions that MOST is based on are largely fulfilled in traditional low resolution ESMs and over homogeneous terrain, and therefore the use of MOST has mostly been successful there. However, increasing body of work has consistently shown the breakdown of ASL scaling over truly complex terrain and more varied, heterogeneous ecosystems where MOST assumptions are violated \cite{de_franceschi_analysis_2009,martins_turbulence_2009,nadeau_similarity_2013,sfyri_scalar-flux_2018,stiperski_scaling_2019}. At the same time, increasingly high resolution numerical models (both for operational weather forecast as well as research Large Eddy Simulations) that seek to capture surface heterogeneity and other complexities, as well as subgrid atmospheric motion highlight the deficiencies of the traditional MOST schemes and the corrections developed until the present \citep{panofsky_characteristics_1977,yaglom_fluctuation_1994,zilitinkevich_influence_2006,nieuwstadt_turbulent_1984,basu_revisiting_2006,zahn_relaxed_2023,arnqvist_fluxprofile_2015}. As models have veered towards configurations that deviate more and more significantly from the embedded MOST assumptions, MOST inaccuracies have become significant bottlenecks in the representation of land-atmosphere interactions.

\par MOST states that for planar homogeneous flow statistics, under stationary conditions with zero mean vertical velocity, turbulence is fully determined by the surface buoyancy flux and momentum flux, as well as the height above the surface. From these quantities, a non-dimensional stability parameter $\zeta$ can be formed, comprised of two length-scales. The first length scale is height $z = z_h - d$, where $z_h$ is the height above ground and $d$ is the zero plane displacement height, negligible over short vegetation. The second is the Obukhov length $L$, defined as $L=-u_*^3\overline{\theta_v}/\kappa g \overline{w'\theta_v'}$, with $\theta_v$ as the virtual potential temperature, $g$ as the acceleration due to gravity, $\kappa=.4$ as the von Kármán constant, $\overline{w'\theta_v'}$ as the buoyancy flux, and friction velocity $ u_*=(\overline{u'w'}^2+\overline{v'w'}^2)^{1/4}$. The ratio of $z$ and $L$ leads to $\zeta = z/L$, a measure of stability (indicating the importance of buoyancy over shear), and the appropriate dimensionless quantity to describe flow statistics in the ASL \citep{monin_basic_nodate}. The theory states that any mean quantity $x$, when non-dimensionalized appropriately by some turbulent scaling variable $x_*$, will be a universal function of only $\zeta$, or: 
\begin{equation}
	\Phi_x(\zeta)=\frac{x}{x_*}.
\end{equation}
The forms of the scaling relations $\Phi_x$ have been obtained by curve fitting through observational results, with the resulting curves applied widely in surface layer parameterization schemes. 

\par The limitations of MOST, however, are apparent in a variety of circumstances, especially as it is frequently applied outside of the atmospheric and land surface regimes where it was originally developed, and where MOST key assumptions are known to not apply. The scaling of velocity variables is well studied and shows some consistent behavior, including consistent deviations from MOST. Under unstable atmospheric stratification, the horizontal variances often fail to scale (data show large scatter) even over horizontally homogeneous and flat terrain \citep{wyngaard_evolution_1974,kader_mean_1990,stiperski_2018}. Vertical velocity also fails to scale under stable conditions \citep{mahrt_stratified_1999}. More generally, in the near neutral regime, as the buoyancy flux goes to zero, the temperature variance can remain finite due to small scale variability in surface temperatures which can be exacerbated by non-stationarity in the time series \citep{chor_flux-variance_2017,sfyri_scalar-flux_2018,Kroon1995,Wyngaard1971,waterman_examining_2022} resulting in deviations of data from the MOST predicted neutral limit. Intermittent turbulence, especially that caused by sub-mesoscale motions and small secondary circulations, violates the assumptions of stationarity and causes a breakdown of many relations, including temperature and velocity variance scaling which can be over-predicted without proper consideration for this intermittence. \citep{falocchi_refinement_2018,mahrt_stratified_1999,lee_influence_2009,waterman_examining_2022}. Broadly, the scaling of various turbulent quantities fails over heterogeneous landscapes with complex terrain where MOST assumptions are rarely satisfied, including high topographic gradients, sparse forest structures, or highly varied land cover \citep{cellier_flux-gradient_1992,moraes_nocturnal_2004,park_effects_2006,de_franceschi_analysis_2009,martins_turbulence_2009,nadeau_similarity_2013,babic_evaluation_2016,babic_fluxvariance_2016,rotach_investigating_2017,finnigan_boundary-layer_2020,kroon_crau_1995,detto_surface_2008,zhou_effects_2019,de_roo_influence_2018}.

\par Despite the widespread issues for the scaling of velocity variances under MOST in complex conditions, it is routinely applied in over these conditions in the overwhelming majority of ASL schemes found in weather and climate prediction models. Many modern ABL schemes use higher order closure (1.5 or 2nd order) schemes \citep{cohen_review_2015,larson_clubb-silhs_2022}. These schemes require information about the variances (velocity for 1.5 order schemes, as well as temperature and moisture for 2nd order schemes) transferred through the ASL in addition to the means. For this information, models rely on MOST scaling. Alternatives to traditional formulations have been proposed to account for these deficiencies with mixed success. Many face limitations in certain atmospheric regimes or, similar to traditional MOST, are not intended to be applied in complex landscapes, \citep{panofsky_characteristics_1977,yaglom_fluctuation_1994,zilitinkevich_influence_2006,nieuwstadt_turbulent_1984,basu_revisiting_2006,zahn_relaxed_2023}, are limited to specific types of scaling \citep{panofsky_characteristics_1977,yaglom_fluctuation_1994,zilitinkevich_influence_2006}, or rely on variables that are hard to measure and model \citep{arnqvist_fluxprofile_2015}. At present, most of these corrections and alternatives have not reached broad use in model parameterizations.  As a result, alternatives to traditional MOST have been an active area of research for nearly five decades with universal agreement on the need for new approaches, especially with increasing model resolution \citep{foken_50_2006}.

\par Recent work by \citet{stiperski_generalizing_2023}, henceforth referred to as SC23, as well as  \citet{mosso_fluxgradient_2024} and \citet{charrondiere2024}, lays out a proposed generalization of MOST that leverages turbulence anisotropy and applies over all complex surfaces and thermal stratifications for scaling of velocity variances \citep{stiperski_generalizing_2023}, gradients \citep{mosso_fluxgradient_2024} and spectra \citep{charrondiere2024}. The results from these works show an ability to account for some of the deviations from, and limitations of, MOST scaling under stable, neutral and unstable atmospheric stratification. These works also found that the new scaling formulations were quite robust, and applied over a wide variety of topographies, although testing over landscapes with significant vegetated canopies, coastal areas, and other ecosystems has yet to be explored. In this follow-up study, we leverage the 47 site National Ecological Observation Network (NEON) to examine the scaling of the variances of streamwise, spanwise and vertical velocity across a wide variety of ecosystems, evaluate the performance of the anisotropy based generalizations of MOST in SC23, and extend our understanding of MOST across various forms and degrees of turbulence anisotropy, with the goal of bringing the SC23 relations closer to future model application for improved ASL schemes.

\par The paper is organized as follows.  In Section \ref{sec:data} we first outline the Data and Methods used in this work, including additional background on the anisotropy of turbulence, details on the SC23 anisotropy-generalized MOST, the NEON data, turbulence data processing, and curve fitting algorithm used. Section \ref{sec:results} shows the Results, including the relations observed in the data, the performance of the SC23 curves as well as refit curves, the variability in behavior across degrees and forms of turbulence anisotropy, as well as inter-site variability. Finally, Section \ref{sec:discussion} brings a discussion of additional points of interest, such as the role of canopy in scaling, the one-component vs two-component turbulence anisotropy, as well as issues of self-correlation. 


\section{Data and Methods}
\label{sec:data}
\subsection{Generalized MOST}
ASL turbulence is anisotropic in eddy sizes and in the kinetic energy content carried by each velocity component.  Such anisotropy is due to differences between characteristic horizontal and vertical length scales driving the turbulent flow, as well as primary mechanisms for turbulence production and removal being directional (i.e. horizontal shear, vertical buoyancy, wall blocking). Including anisotropy may be able to indirectly account for some processes, particularly turbulent transport, which are neglected in traditional theories. As common in studies of the anisotropy of the Reynolds stress tensor, $\overline{u_i'u_j'}$, we decompose it into isotropic and anisotropic components, with the normalized anisotropy tensor in non-dimensional form written as: $b_{ij}=\frac{\overline{u_i'u_j'}}{\overline{u_l'u_l'}}-\frac{1}{3}\delta_{ij}$ \citep{Pope_2000}, where the indices $i$ and $j$ vary from 1 to 3 representing the three directions in the Cartesian coordinate system, and index repetition implies sum over the components. The $\overline{u_lu_l}$ is twice the turbulence kinetic energy and $\delta$ is the Kronecher delta. This symmetric tensor, has (infinite sets of) three invariants, the first of which is null by definition, and the other two, following the work of \cite{lumley_newman_1977}, contain the full information on the stress tensor and allow us to examine typical behavior within the invariant map known as the Lumley triangle. Further, to overcome the nonlinear nature of the mapping proposed by \cite{lumley_newman_1977}, we follow the work of \citet{banerjee_presentation_2007}, which utilizes an alternative set of invariants offer a linear representation known as the barycentric map of the Lumley triangle. In this approach from \citet{banerjee_presentation_2007}, the invariant $y_b$ is defined as a function of the smallest eigenvalue of the anisotropy tensor, and represents the degree of anisotropy, such that
\begin{equation}
    y_b=\sqrt{3}/2(3\lambda_3+1)
    \label{yb}
\end{equation}
The second invariant $x_b$ is related to the form of anisotropy (one or two-component), and is written as
\begin{equation}
    x_b=\lambda_1-\lambda_2+\frac{1}{2}(3\lambda_3+1)
    \label{xb}
\end{equation}
where $\lambda_1$, $\lambda_2$, $\lambda_3$ are the eigenvalues of the normalized anisotropy tensor $b_{ij}$ sorted from largest to smallest. In the works of SC23, the invariant $x_b$ was found to be less significant for MOST scaling \citep{stiperski_generalizing_2023}. When $y_b=0$, it implies highly anisotropic turbulence in one or two components, and $y_b=\sqrt{3}/2$ for purely isotropic turbulence. When $x_b$ is small, this signifies that the turbulence is primarily two-component (i.e. two of the eigenvalues similarly dominate) and high values of $x_b$ are associated with one component turbulence anisotropy where a single eigenvalue dominates over the others. 

\par In SC23 $y_b$ is used to extend MOST scaling for the velocity variances, based on results showing that by binning data into regimes according to the degree of anisotropy, distinct curves for each regime emerge \citep{stiperski_generalizing_2023}. Including $y_b$ may improve MOST by indirectly accounting for some of the (transport) processes neglected by the traditional simplifying assumptions. One common set of traditional MOST relations \citep{stiperski_scaling_2019,panofsky_characteristics_1977,Stull1994,kaimal_atmospheric_1994} has the following forms:
\begin{equation}
    \Phi_x=\sigma_x/u_*=a_x(1-3\zeta)^{1/3} \ \ \ \ \ \ \ (\zeta<0).
    \label{trad_u}
\end{equation}
With $a_u=2.55$, $a_v=2.05$ and $a_w=1.35$, and for stable conditions 
\begin{equation}
    \Phi_x=\sigma_x/u_*=a_x \ \ \ \ \ \ \ (\zeta>0)
    \label{trad_s}
\end{equation}
with $a_{u,v}=2.06$ and $a_w=1.6$, as in SC23. The anisotropy can then modify these traditionally constant scaling curve parameters, $a_x$ to be variable functions of $y_b$. This yields new forms for scaling relations:
\begin{equation}
    \Phi_x=\sigma_x/u_*=a_x(y_b)[1-3\zeta]^{1/3} \ \ \ \ \ \ \ (\zeta<0).
    \label{sc23_u}
\end{equation}
and 
\begin{equation}
    \Phi_x=\sigma_x/u_*=a_x(y_b)[1+3\zeta]^{d_x(y_b)} \ \ \ \ \ \ \ (\zeta>0)
    \label{sc23_s}
\end{equation}
where the parameters $a_x$ and $d_x$ are no longer constant but functions of $y_b$.
Additional details can be found in SC23 and its supplementary material \citep{stiperski_supplemental_nodate}. The modification of existing MOST formulations allows for a seamless transition in various boundary layer schemes which utilize MOST scaling. The shared form between the two means that they retain widely accepted theoretical limits, including the 1/3 power law under free convection and the z-less limit in stable stratification.

\subsection{Data: National Ecological Observation Network}
\begin{figure}
\centering
\includegraphics[width=6in]{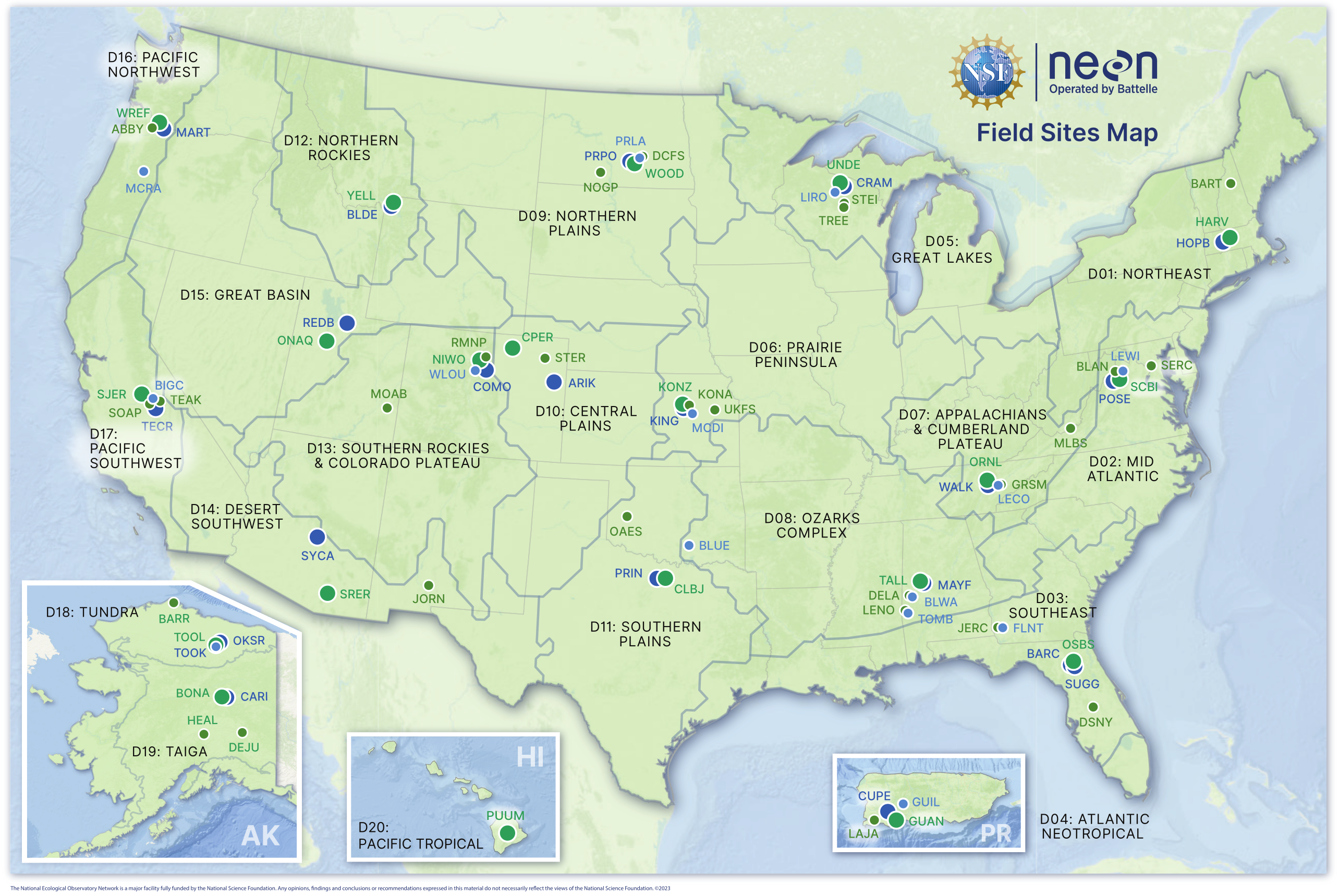}
\caption{NEON site map adapted from NEON resources. All sites in green contain sonic anemometers necessary for the analysis in this manuscript; blue sites are aquatic and not represented in this study. Larger circles represent the primary site(s) for each given ecological domain. NEON (National Ecological Observatory Network). Field Sites Map - Poster w Site Index. https://www.neonscience.org/sites/default/files/FieldSitesMap-33x18-PosterwIndex.pdf (accessed 17 Mar 2024)}
\label{fig:neon_map}
\end{figure}
NEON is an NSF funded network with sites placed in ecological domains and deliberately selected to represent a broad cross section of landscapes and ecosystems within United States territory, including Alaska, Hawaii and Puerto Rico as seen in figure \ref{fig:neon_map}. All sites use the same instrumentation and methodology, and have well documented, publicly available data products. Depending on the site, NEON has published meteorological tower data from as early as 2016 through to the present day, with an average record of 8 years (as of December 2024). We use all data regardless of season. Towers at sites with a canopy less than three meters are designed to be 8m tall, whereas towers at sites with a canopy greater than three meters are designed to have a height corresponding to $d+4(h_c-d)$, with canopy height $h_c$ and zero plane displacement height $d$, to ensure that the turbulence exchange assembly samples turbulence largely above the momentum roughness layer \citep{Metzger2019}. A sonic anemometer at the top collects 20 Hz measurements of velocity and sonic temperature, co-located with a gas analyzer to collect 20 Hz CO2 and H2O concentrations. NEON provides 30 minute averages of the fluxes and variances of these quantities, as well the raw sensor output and modeled flux footprint shape to relate point measurements to the area perceived by the sensors \citep{national_ecological_observatory_network_neon_bundled_2023}. NEON also provides high resolution imagery from a tower mounted camera at 15 minute intervals, which are further analyzed to provide products including using changes in the green chromatic coordinate of the imagery to identify the local growing season \citep{richardson_tracking_2018}. Approximately once per year during the summer, NEON conducts airborne remote sensing flights over each terrestrial site, covering an area of approximately 12 km by 12 km with 1 meter resolution LiDAR and spectral imaging. National Land Cover Database (NLCD), a Landsat derived product defining the land cover classification at a high (30m) resolution over CONUS for 2021, and Alaska in 2016, has been co-located with each of the NEON domains \citep{jin_overall_2019}. 
\par The NEON airborne datasets are combined with the flux footprint information and the NLCD to determine the dominant land cover as well as metrics of surface complexity within area perceived by the 3D sonic anemometer. Surface complexity is challenging to quantify, with no definitive metric that can completely express how the topography and canopy structure can impact turbulence. To provide a simple, broad sense of complexity at each site, we compute the spatial standard deviation (i.e. spatial variability) of three metrics within the flux footprint: canopy height ($\sigma_{hc}$) to capture vegetative complexity, the Digital Terrain Model (DTM) ($\sigma_{DTM}$) to capture terrain complexity, and the Digital Surface Model (DSM) ($\sigma_{DSM}$) as a metric of surface complexity that includes both topography and canopy complexity. In table \ref{tab:sites} we show the values of these metrics across the sites, illustrating the wide range of complexity across NEON. Sites are categorized into Less Complex ($\sigma_{DSM}\leq10$), Complex ($10<\sigma_{DSM}\leq20$) and Very Complex ($\sigma_{DSM}>20$).

\begin{table}[t]
\caption{Table Showing the NEON sites, their dominant NLCD land cover and complexity. Three complexity metrics are shown for each site: median spatial standard deviation of the digital surface model (DSM) within the tower flux footprint, median spatial standard deviation of the digital terrain model (DTM) within the tower flux footprint, and median spatial standard deviation of the canopy height within the tower flux footprint. Abbreviated land cover types are: AK:se for Alaskan Sedge, DwScr for Alaskan Dwarf Scrub, Wtlnd for Wetland, Evrgn for Evergreen, Decid for Deciduous, and MixFr for Mixed Forest.}
\small
	\label{tab:sites}
	\begin{center}
    \setlength{\tabcolsep}{3pt}
    \renewcommand{\arraystretch}{.9}
	\begin{tabular}{l l ccc | l l ccc | l l ccc }
			\hline
			   \multicolumn{5}{c}{\textbf{Less Complex}} & \multicolumn{5}{c}{\textbf{Complex}} & \multicolumn{5}{c}{\textbf{Very Complex}} \\ \hline
              Site & NLCD & $\sigma_{DSM}$ & $\sigma_{DTM}$ & $\sigma_{hc}$  & Site & NLCD & $\sigma_{DSM}$ & $\sigma_{DTM}$ & $\sigma_{hc}$  & Site & NLCD & $\sigma_{DSM}$ & $\sigma_{DTM}$ & $\sigma_{hc}$  \\ \hline 
              \textbf{BARR} & AK:Se & $<1$&$<1$&$<1$& \textbf{UNDE}& Wtlnd & 11  & 6   & 8     & \textbf{HARV} & MixFr & 21  & 19  & 9 \\
              \textbf{JORN} & Shrub & 2 & 2   &$<1$   &  \textbf{PUUM} & Evrgn & 14  & 12  & 5     & \textbf{SJER} & Shrub & 31  & 31  & 3 \\
              \textbf{KONA} & Crops & 2  & 2  & $<1$    & \textbf{STEI} & Wtlnd & 14  & 10  & 7    & \textbf{GRSM} & MixFr & 36  & 38  & 9 \\
              \textbf{KONZ} & Grass & 2   & 2  & $<1$    &  \textbf{TREE} & MixFr & 14  & 10  & 8     & \textbf{BONA} & Wtlnd & 36  & 37  & 2 \\
              \textbf{STER} & Crops & 2   & 2   &$<1$   &  \textbf{SERC} & Decid & 15  & 9   & 12    & \textbf{SCBI} & Decid & 41  & 38  & 10 \\
              \textbf{CPER} & Grass & 3   & 3   & $<1$  &  \textbf{MLBS} & Decid & 17  & 17  & 5     & \textbf{GUAN} & Evrgn & 44  & 44  & 3 \\
              \textbf{SRER} & Shrub & 3   & 3   & 1     &  \textbf{ORNL} & Decid & 17  & 13  & 8     & \textbf{NIWO} & Grass & 51  & 52  & 3 \\
              \textbf{LAJA} & Crops & 3   & 3  & 1    &  \textbf{YELL }& Shrub & 18  & 18  & 2     & \textbf{BART} & MixFr & 59  & 59  & 5 \\
              \textbf{OAES} & Grass & 4   & 4   & 1     &  \textbf{UKFS} & Decid & 19  & 17  & 5     & \textbf{WREF} & Evrgn & 99  & 99  & 11 \\
              \textbf{WOOD} & Grass & 4   & 4   & $<1$ &  \textbf{ABBY} & Evrgn & 19  & 19  & 13    & \textbf{TEAK} & Evrgn & 117 & 116 & 7  \\
              \textbf{TOOL} & DwScr & 4   & 4   & $<1$  & \textbf{TALL} & Evrgn & 20  & 20  & 10    & \textbf{SOAP} & Grass & 128 & 127 & 7  \\
              \textbf{ONAQ} & Shrub & 5   & 5   & $<1$  & & & & & & \textbf{RMNP} & Evrgn & 162 & 163 & 5   \\
              \textbf{DEJU} & Evrgn & 5   & 2   & 4     & & & & & & & & & &  \\
              \textbf{DSNY} & Pastr & 5   & 0   & 5     & & & & & & & & & & \\
              \textbf{MOAB} & Shrub & 6   & 6   & $<1$  & & & & & & & & & & \\
              \textbf{DCFS} & Grass & 6   & 6   & $<1$  & & & & & & & & & &  \\
              \textbf{NOGP} & Grass & 6   & 6   & 2     & & & & & & & & & & \\
              \textbf{HEAL} & Shrub & 7   & 7   & 1     & & & & & & & & & & \\
              \textbf{BLAN} & Pastr & 8   & 5   & 6     & & & & & & & & & & \\
              \textbf{DELA} & Wtlnd & 9   & 4   & 8     & & & & & & & & & & \\
              \textbf{CLBJ} & Decid & 9   & 9   & 4     & & & & & & & & & & \\
              \textbf{JERC} & Evrgn & 9   & 4   & 9     & & & & & & & & & & \\
              \textbf{LENO} & Wtlnd & 9   & 7   & 8     & & & & & & & & & & \\
              \textbf{OSBS} & Evrgn & 10  & 7   & 6     & & & & & & & & & & \\
               
		\end{tabular}
	\end{center}
\end{table}

\subsection{Data Processing}
Most of the data used as part of this work relies on publicly available post-processed products from NEON. The raw turbulence data was despiked, and a planar-fit coordinate rotation was performed for each day with a 9 day moving window. Lag correction is then performed, statistics are computed for the 5 minute and 30 minute averaging periods, and a wavelet-based high-frequency spectral correction \citep{nordbo_wavelet-based_2013} is performed for the fluxes. More details of the post-processing can be found in the NEON turbulence algorithm theoretical basis document \citep{metzger_neon_2022}. The selection of averaging period is critically important to adequately separate the turbulence from non-turbulent motions and because anisotropy is a scale dependent phenomena, as most larger scale phenomenon are highly anisotropic. During stable stratification, we selected a 5 minute averaging period and a 30 minute averaging period for unstable stratification. A 5 minute averaging period was chosen after conducting spectral analysis (multi-resolution decomposition (MRD), \citep{vickers_cospectral_2003,howell_multiresolution_1997,stiperski_supplemental_nodate}) to determine the location of the spectral gap and point where $\overline{w'\theta'}$ scale-wise contribution switches sign across a representative subset of the NEON sites as in \cite{stiperski_supplemental_nodate}. Details of this analysis can be found in supporting information S1 and S2. The 5 minute averaging period was chosen, in contrast to the traditional 30 minute averaging period or the 1 minute averaging period used in SC23, to balance capturing the majority of the spectra against excluding highly anisotropic and potentially non-turbulent large scale processes. The MRD analysis suggest that 30 minute averaging period would have likely led to significant non-local contributions and erroneous classifications of anisotropy, and a one minute period only captured a sufficient portion of the spectra for sites without forests (similar to those in SC23).
\par The 5-minute fluxes are unavailable from NEON, as are the 5 minute and 30 minute Reynolds stresses needed to assess anisotropy. As such, these quantities were computed directly by the authors from the raw turbulence statistics. Processing of the 5 minute quantities followed the NEON processing algorithm using a combination of author-written python scripts and the NEON managed eddy4R package. The 5 minute fluxes do not include the spectral correction, which appeared to have a minimal impact on the processing. After all processing, we use quality flags from NEON to quality check the data. NEON quality checks include checking for stationarity by comparing the original and detrended timeseries for the first moments \citep{vickers_quality_1997} and by comparing the 30 minute averaging period covariance to the covariances of the six, 5 minute averaging periods \citep{foken_tools_1996}. For the 5 minute fluxes and variances, we instead use 5, 1 minute averaging periods to assess stationarity. In addition, statistical plausibility tests, a test to check to see if the turbulence is well developed, and instrument quality flags are applied. For more details on NEON quality checks and algorithmic implementation, see pages 28-33 of \citep{metzger_neon_2022}. 
\par After all quality checks, approximately 4.5 million points remain under stable stratification and 1 million points under unstable stratification for the velocities, corresponding to approximately one site-year for each stratification. Data availability varies between sites, with site availability at most sites ranging between 75\% and 125\% of the median site (median availability is approximately 20,000 points under unstable conditions and 100,000 points under stable conditions). Available data varies across land cover, with 21\% of the data at Evergreen sites, 16\% Grassland, 15\% Deciduous, 13\% Shrub, 12\% Wetland, 10\% Mixed Forest, 6\% Cropland, 5\% Pasture, and less than 1\% Dwarf Scrub and Alaskan Sedge. In total, this represents a fairly even split for the data between forested (46\%) and non-forested (42\%) ecosystems, with wetlands, most of which are forested, constituting the remaining $12\%$. The data availability in the stability-anisotropy plane is shown in figure S3. The majority of the data is concentrated around stabilities ranging from $|\zeta|=10^1$ to $|\zeta|=10^{-2}$ and with $y_b$ between $0.15$ and $0.5$, and $x_b$ between 0.2 and 0.6. Notably absent is highly isotropic (and, to a lesser extent, highly anisotropic) turbulence and one-component turbulence anisotropy under near neutral stratification.

\subsection{Curve Fitting Methodology}
While SC23 does provide the forms of the functions for the various parameters $a(y_b)$, $d(y_b)$ etc., and we evaluate them with their original values in this work, it will be useful to re-fit them to this larger dataset. We use a two step curve fitting process to determine functions for the equation parameters as a function of anisotropy $y_b$. All data is combined into one dataset for analysis. For each variable, and for each two groupings of stratification, $\Phi$ is divided according to $y_b$ into 100 bins with equal numbers of points in each bin. For each bin, the functional forms in equations \eqref{sc23_u} and \eqref{sc23_s} are fit to the data, treating $a_x$ and $d_x$ as constant parameters to fit to. The curve fitting uses a Cauchy loss function, which is generally considered robust to noise. This produces a series of best fit parameter values for $a_x$ and $d_x$ that vary across the bins of anisotropy. The parameter-anisotropy curves produced by the first curve fitting are then fit themselves to a polynomial function (varying from degree 1 to 3). For unstable stratification, this fit is conducted on $\log_{10}y_b$ instead of on $y_b$, as was done for unstable conditions in \cite{stiperski_generalizing_2023}. The refit often requires fewer parameters than SC23 (with lower degree polynomials) and in no cases requires more than SC23. The parameter values are shown in table S1 in the supporting information.


\section{Results}
\label{sec:results}


\subsection{Observed Dependence of Scaling Relations on Anisotropy}
\begin{figure}
\centering
\includegraphics[width=5in]{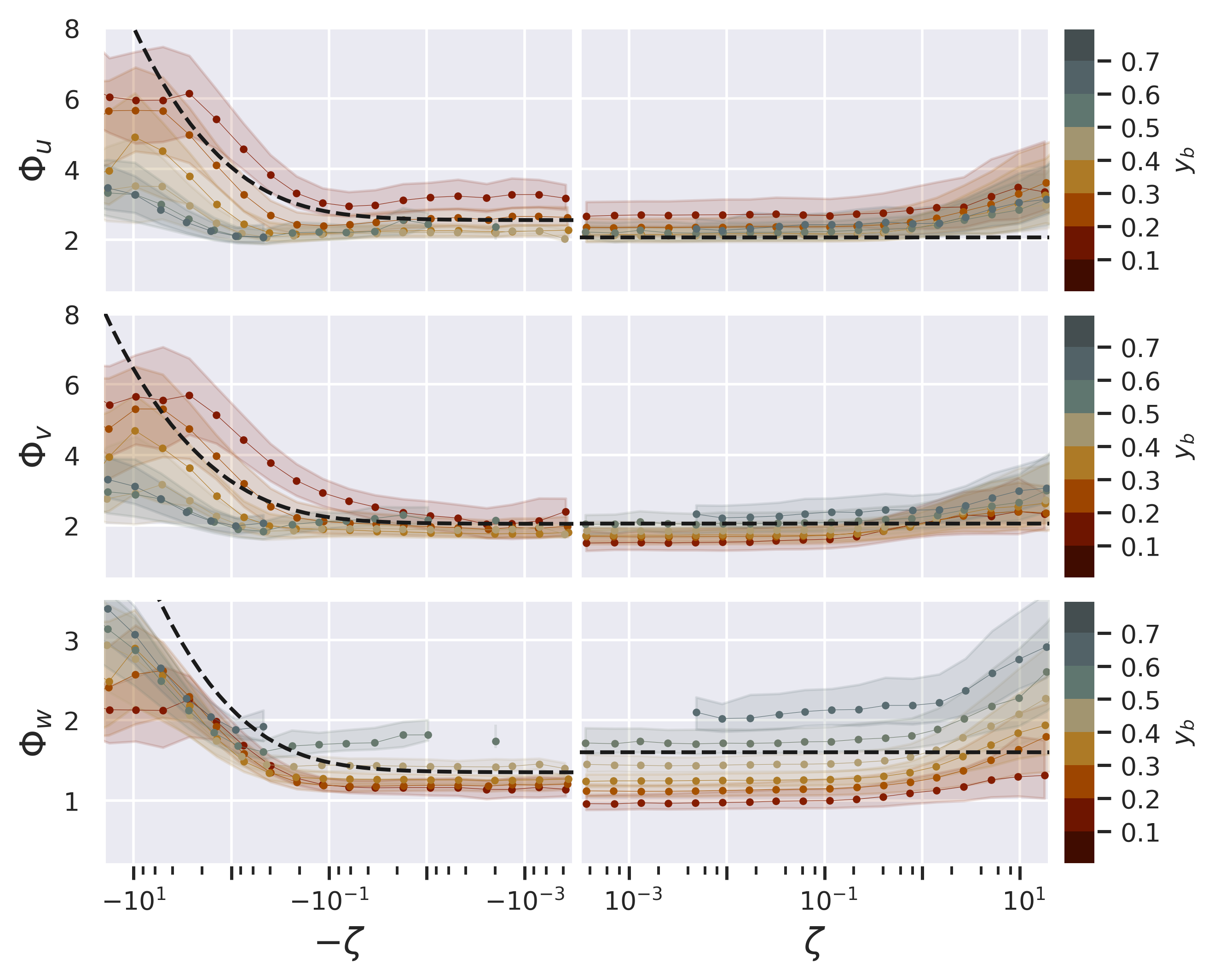}
\caption{Observed scaling for the square root of the variance of, from top to bottom, non-dimensionalized streamwise velocity $\Phi_u$, spanwise velocity $\Phi_v$ and vertical velocity $\Phi_w$. Scatterplots show 100,000 random points across sites, colored by $y_b$ with red being anisotropic and blue being isotropic. Lines are colored by $y_b$ with red being anisotropic and blue being isotropic and show the median values for $\Phi$ across linearly spaced bins of anisotropy and logarithmically spaced bins of $\zeta$. The area between the 1st and 3rd quartile for each bin is shaded. Black dashed line shows a traditional MOST scaling (equations \eqref{trad_u},\eqref{trad_s}) for these non-dimensionalized quantities.}
\label{fig:basic_scaling}
\end{figure}

A basic examination of how the flux-variance relations for velocity variances vary with stability and anisotropy shows a strong dependency on $y_b$ in median scaling curves for many variables (figure \ref{fig:basic_scaling}). In the scaling of the vertical velocity under stable stratification, in particular, there is a clear separation of the data according to anisotropy that covers nearly the full observed range of scatter in turbulence statistics.  An especially strong dependence on $y_b$ is clear for streamwise and spanwise velocity under unstable stratification as well, implying that anisotropic turbulence carries more overall turbulent kinetic energy (TKE) for the same given stratification. This may imply a connection between anisotropy and the boundary layer depth, as previous work shows a strong connection between the scaling of the horizontal velocities under convective conditions and boundary layer depth, however, given the limited large scale data available from NEON, we reserve a more in depth evaluation of this phenomena for future work with more appropriate datasets. As we approach free convective conditions ($\zeta < -1$), nearly all quantities in the binned data lose an obvious dependence on $\zeta$ (curves flatten out), deviating from both traditional scaling and the SC23 curves which show continued $\zeta$ dependence with the typical $1/3$ power law. This is particularly true for more anisotropic turbulence. It is worth noting that data availability in this region of $y_b$-$\zeta$ space is poor when compared to other regions. This makes it more difficult to assess if this is an observation due to data quality and sampling issues, or a true aspect of scaling as we approach free convection. In addition $u_*$ is quite small in this regime compared to the buoyancy flux, making the computed scaling terms quite sensitive to small changes in $u_*$ which may impact the results especially given a limited sampling size. The $1/3$ power law limit under free convection ($\zeta>1$) is relatively well established for varied conditions with both field data and theory supporting \citep{Wyngaard1971,Nadeau2013,rannik_surface_1998,kroon_crau_1995,detto_surface_2008,tong_moninobukhov_2018}, so this is most likely an error. Prior to applying quality assurance as specified in section \ref{sec:data}, the traditional limit here was followed more closely. A possible explanation for this discrepancy is that the quality check for well developed turbulence eliminates $\Phi$ values with low $u_*$ in this regime, where low shear is common and can lead to high $\Phi$ values, but potentially maintains the more anomolous data with higher $u_*$ values. Another notable difference is a clear dependence on anisotropy for spanwise variance under unstable stratification that was not reported in SC23. Under stable stratification, there is a weak relationship (aside from $\Phi_w$, where it is strong), between anisotropy and the scaled variables in accordance with previous work. $\Phi_u$ and $\Phi_v$ appear to have equal magnitude (but opposite) scaling with $y_b$, whereas $\Phi_w$ scales strongly positively with $y_b$, indicating that, in sum, more isotropic turbulence carries slightly higher $TKE$ for the same level of stratification, an inverse of the relationship under unstable stratification. 

\subsection{Performance of Generalized MOST SC23 Curves}
\begin{figure}
\centering
\includegraphics[width=5in]{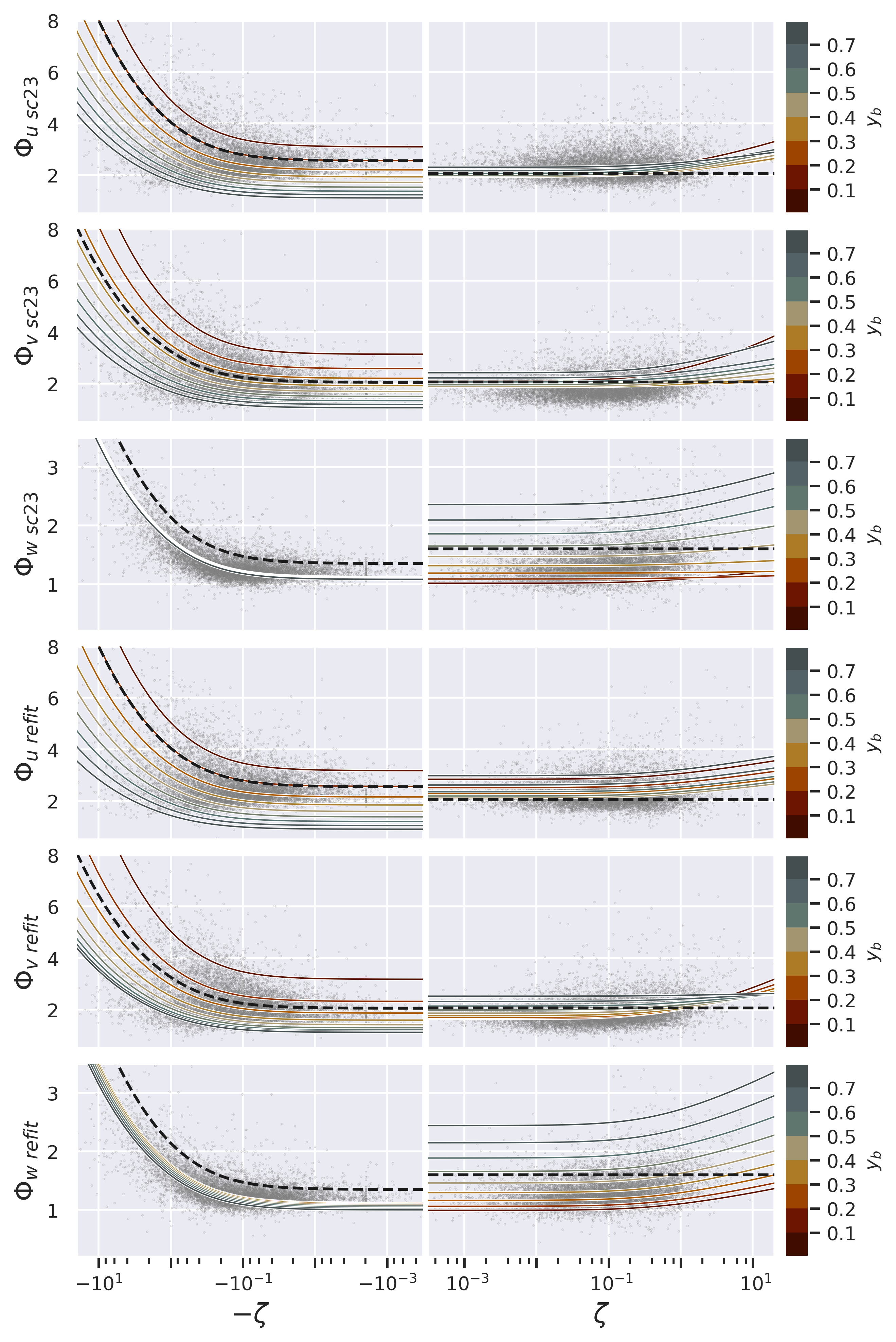}
\caption{Scaling for the square root of the variance of, from top to bottom, non-dimensionalized streamwise velocity $\Phi_u$, spanwise velocity $\Phi_v$ and vertical velocity $\Phi_w$ from SC23 followed by a refit of these relations to the NEON data (equations \eqref{sc23_s},\eqref{sc23_u}). Scatter shows 10,000 randomly selected points for each scaling relation. Lines show the scaling relations by anisotropy for 9 evenly spaced $y_b$ levels between 0.1 and 0.7. Black dashed line shows a traditional MOST scaling (equations \eqref{trad_s},\eqref{trad_u}) for these non-dimensionalized quantities.}
\label{fig:stip_scaling}
\end{figure}

For the scaling of variances in the top three panels of figure \ref{fig:stip_scaling}, the SC23 curves appear to visually cover the range of the data and are very similar to the median fits seen in figure \ref{fig:basic_scaling}. Some small differences are worth noting. Particularly, there seems to be a stronger separation by anisotropy in the streamwise and vertical velocity under stable conditions in the data than in the SC23 curves. Another point of difference between the data and the SC23 curves is found under very unstable stratification as we approach free convective conditions, likely due to data issues in this regime as discussed previously.

The SC23 relations were not originally fit over terrain with tall vegetation or over some less commonly explored ecosystems (boreal north, tundra, tropical, coastal), which could at least partially explain the differences over the NEON sites. Fitting similar to that conducted in the previous work is done over the NEON sites to re-fit the parameters for this network, to suggest any modifications or improvements, and to evaluate the robustness of the SC23 relations. The mathematical representation of the refit and the SC23 relations remains the same and is shown in equations \eqref{sc23_u} and \eqref{sc23_s}, the functions for parameters $a_x(y_b)$ and $d_x(y_b)$ are shown in the supplemental table S1 and the scaling curves compared to the data are shown in the second set of three panels in figure \ref{fig:stip_scaling}. The curves are generally quite similar in form to SC23, with slightly increased dependence on anisotropy under stable conditions for all three curves, and a downward shift in the curves for $\Phi_v$ scaling under unstable conditions. Also, for $\Phi_u$ and $\Phi_w$ under stable conditions the $d_u$ parameter in equation \eqref{sc23_s} no longer depends on $y_b$ in the refit where $d_u$ was observed to be a constant. The refits are able to yield some performance increases, discussed later in this work, and also a reduction in complexity of the fit when compared to SC23. Under unstable conditions, the degree of the polynomial for $a_w(y_b)$ was reduced by one, and under stable conditions the total number of tunable fit parameters was reduced in half from 8 to 4 for $u$ and $w$ and reduced from 8 to 5 for $v$ with no negative repercussions on fit accuracy. See table S1 for details.

\par To quantitatively evaluate the performance of the anisotropy based generalizations of MOST, we compare them to the traditional MOST relations using a skill score (SS), defined as \citep{stiperski_generalizing_2023}
\begin{equation}
	SS = 1-\frac{MAD_{new}}{MAD_{MOST}},
\end{equation}
where MAD is the median absolute deviation calculated as the median deviation from the scaling curve for each datapoint, or
\begin{equation}
	MAD = \text{median}(|x_{scaling}-x|).
\end{equation}
The skill score has a maximum value of one, indicating the unlikely scenario where the SC23 curves perfectly replicate the observations. Broadly, the skill score can be interpreted as a percent reduction in error. We also examine errors using the median bias, defined as 
\begin{equation}
	Bias = \text{median}(x_{scaling}-x).
\end{equation}
All of these metrics can be used to describe errors and performance differences; generally we utilize $MAD$ and the skill score when we want to directly assess how well a scaling relation can reproduce the observed scaling, and bias is used when we want to understand whether the error from $MAD$ is the result of the scaling over or under predicting.

\begin{figure}
\centering
\includegraphics[width=4in]{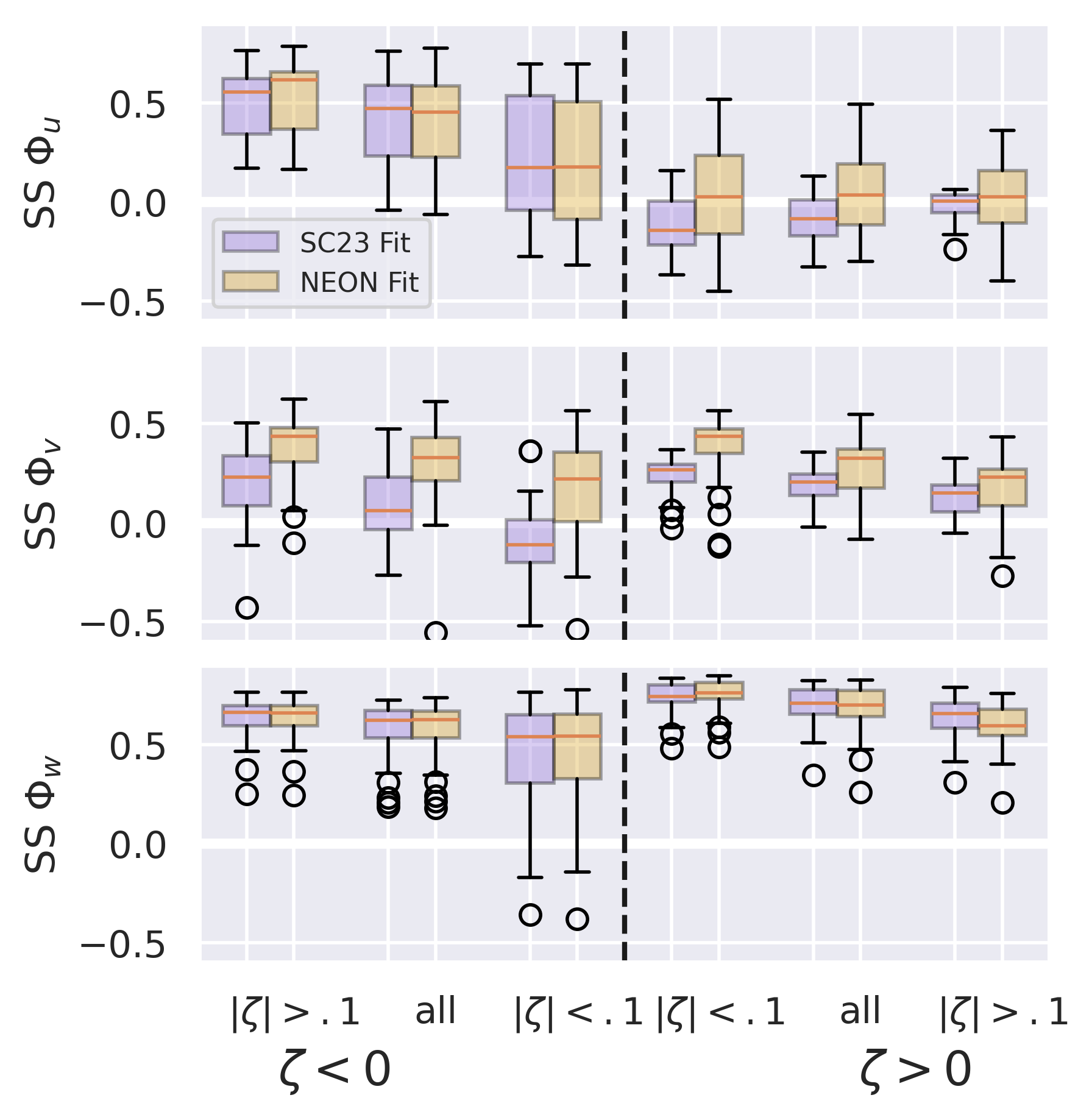}
\caption{Box-whisker plot showing the skill score SS across all 47 sites between the SC23 relations and the traditional MOST scaling (purple) as well as the refit scaling and traditional MOST scaling (orange) for the square root of the variance, from top to bottom, of non-dimensionalized streamwise velocity $\Phi_u$, spanwise velocity $\Phi_v$ and vertical velocity $\Phi_w$. Box-whiskers from left to right show the skill score for different atmospheric stratifications: left to right is SS for more unstable with $|\zeta|>.1$, SS for all unstable stratification, SS for unstable near-neutral stratification ($|\zeta|<.1$), then the stable regieme with SS under stable near-neutral ($|\zeta|<.1$), SS for all stable stratification, and more stable ($|\zeta|>.1$)}
\label{fig:SS_all}
\end{figure}

The skill score of the SC23 relations across the NEON sites (figure \ref{fig:SS_all}) is strong and largely consistent with the skill score found over the original, non-NEON (non-vegetated) sites evaluated previously. It shows that, as an aggregate across sites, the relations reduce scatter and broadly outperform MOST. The best performance comes in the scaling of vertical velocity in stable stratification, where SC23 represents an improvement compared to traditional scaling at every single one of the 47 sites for all stability classes in those ranges and median improvement near 75\%. For the scaling of $\Phi_u$ under unstable conditions as well there is a consistent observed improvement in the scaling performance when generalized based on anisotropy, with almost every site seeing a performance improvement and the median site seeing almost a 50\% improvement. 
 
\par In the stable stratification regime, however, the streamwise velocity does not experience the same performance improvement, with only very small positive median SS, and only after refitting. Whereas $\Phi_u$ does show some separation according to anisotropy in figure \ref{fig:basic_scaling}, the SC23 curves have a different dependence on $\zeta$ compared to the NEON observations. This could be responsible for the performance decrease, as could small changes induced by the different averaging periods between the two studies. Regardless, it is worth noting that $\Phi_u$ under stable conditions has the least significant scaling based on $y_b$ in both SC23 and over the NEON sites, implying any performance improvement or observed relationship was likely to be less significant. The SS in spanwise velocity has strong overall performance under unstable stratification, however there is a high variability in SS across sites, which is particularly large under near neutral conditions. The fit curves do not converge to a consistent value across $y_b$ under near neutral conditions, whereas the data appears to show a general convergence which may explain some of the observed variability in SS under near neutral stratification. This weaker performance was also observed in the original SC23 dataset. It is also worth noting that the skill scores for stable conditions may be slightly larger due to the decision to use a constant scaling for the traditional functions in equation \eqref{trad_s}, as opposed to $\zeta$ dependent forms also published in the literature (i.e. \citet{kaimal_atmospheric_1994}) and used as the base for equation \eqref{sc23_s}. The constant form was selected due to both its common application (despite more update forms existing) and to facilitate inter-comparison with the results of the SC23 study.

The SC23 relations perform remarkably well over the NEON tower network without any adjustments or modifications (figure \ref{fig:SS_all}), especially given the fact that they were not developed using any information from vegetated canopies. The skill scores over NEON are generally comparable with those found over the towers originally used to formulate the generalized MOST scaling relations. There is, however, notable drop in performance for SC23 compared to the refit curves when examining the scaling of spanwise velocity $\Phi_v$ across all conditions, as well as $\Phi_u$ under stable conditions.

The refit addresses this performance gap in spanwise velocity scaling, with increases in skill between 10\% and 30\% depending on the regime; this is largely a consequence of a downward shift across all levels of anisotropy where the spanwise velocity scales slightly less with $\zeta$ than over the original towers that lead to the development of the generalized scaling. As detailed in future sections, it appears this is driven by poor performance of the SC23 scaling relations over complex canopies, which the refit successfully corrects. Aside from this rather dramatic difference in scaling of $\Phi_v$, the refit yielded relatively few changes in performance providing further evidence of the robustness of the SC23 relations for streamwise and vertical velocity scaling.

\subsection{Variability by Anisotropy}
\subsubsection{Deviation from Traditional MOST}
\begin{figure}
\centering
\includegraphics[width=4in]{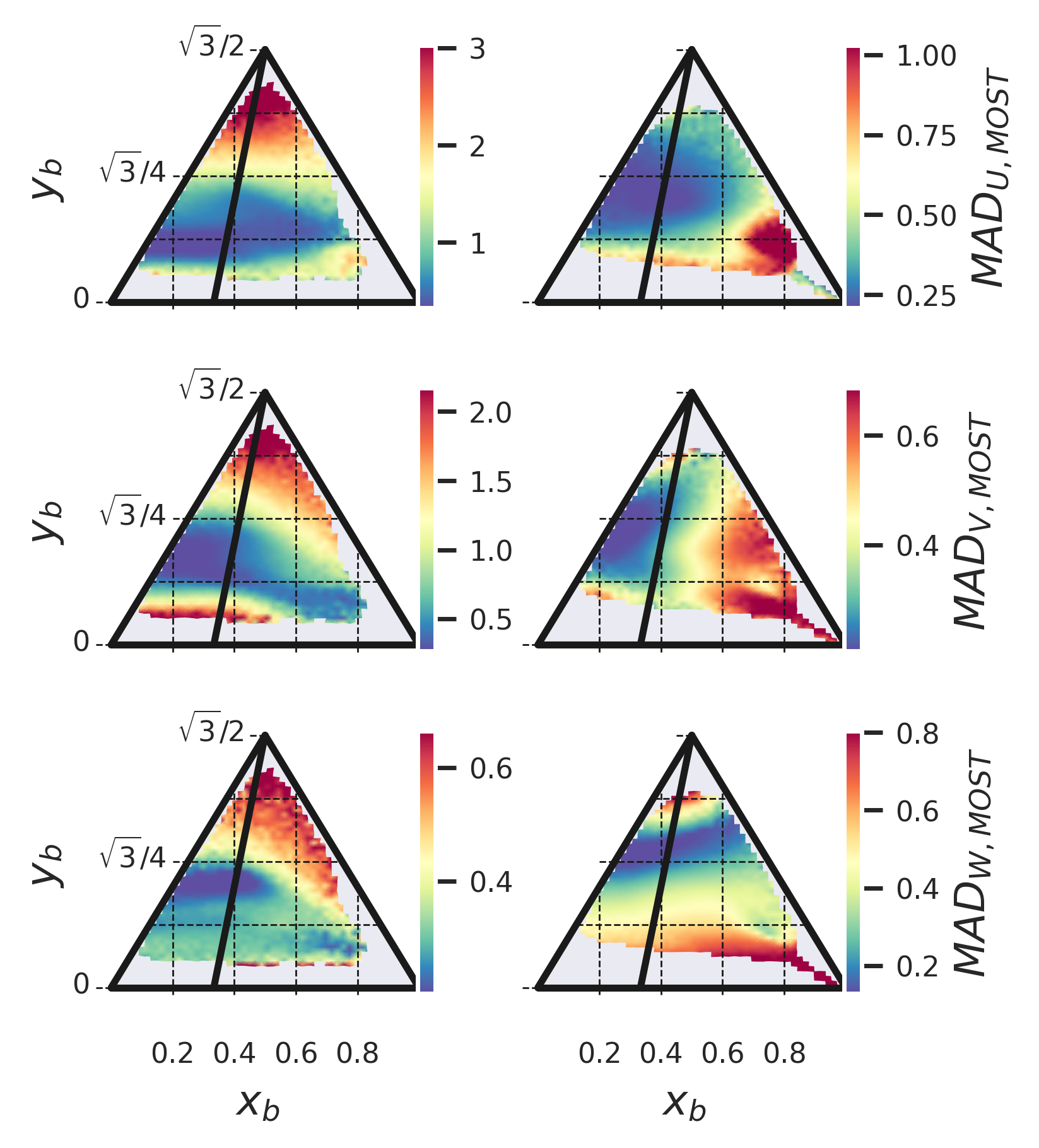}
\caption{Median Absolute Deviation (MAD) of the data from traditional MOST relations mapped onto the Barycentric Map of the Lumley Triangle. Blue signifies lower error and red higher error. MAD shown for unstable (left) and stable (right) conditions for the variances of streamwise (top), spanwise (middle), and vertical (bottom) velocity. The plane-strain line is plotted in black for reference.}
\label{fig:lumley}
\end{figure}

By examining the scaling of velocity variances across anisotropy we can start to understand and evaluate the existing performance gaps and characteristics of the scaling relations. Traditional MOST based scaling appears to perform best under two component (small $x_b$), moderately anisotropic (mid level $y_b$) turbulence as apparent in figure \ref{fig:lumley}. While the exact location of best performance shifts depending on the scaling variable and stability regime, it lies consistently to the left of the plane-strain line near $y_b=\sqrt{3}/4$ implying that traditional MOST assumptions and scaling apply best for turbulence under moderately anisotropic, two component anisotropy. Under unstable conditions, the performance of MOST is largely insensitive to $x_b$ as previously reported \cite{stiperski_anisotropy_2021}. However, for larger values of $x_b$ where turbulence anisotropy is closer to the one component limit, the error gains a slight but significant correlation with $x_b$. This has been previously noted by \cite{stiperski_scaling_2019}, where data with one-component turbulence anisotropy defied scaling. This type of turbulence tends to be encountered in transition periods, or nighttime when trends are large in the data \citep{gucci_sources_2023}. The MAD is larger for very small and very large values of $y_b$ under unstable conditions. This is due to a continuous change in the bias of traditional MOST to the data that $MAD$ removes through the absolute value. MOST overpredicts $\Phi_u$, $\Phi_v$, and $\Phi_w$ for high values of $y_b$ and under-predicts them for low values of $y_b$; details on the bias of MOST in $x_b$-$y_b$ space is found in figure S4. 

Alternatively, for stable conditions deviations from traditional MOST scaling appear to be more equally dependent on $x_b$ and $y_b$. The $x_b$ dependence is most clear in the scaling of $\Phi_u$, where there is very little $y_b$ error dependence under stable stratification. This helps to explain the relatively poor performance of SC23 in this regime seen in figure \ref{fig:SS_all}. The scaling of the spanwise velocity shows a more mixed dependence on both $y_b$ and $x_b$, with the performance gradient approximately parallel to the plane-strain line, although once again $x_b$ boasts the stronger relationship. For vertical velocity variance scaling under stable conditions, the error largely changes with $y_b$ as in the unstable scaling for the other variances.

\begin{figure}
\centering
\includegraphics[width=5.5in]{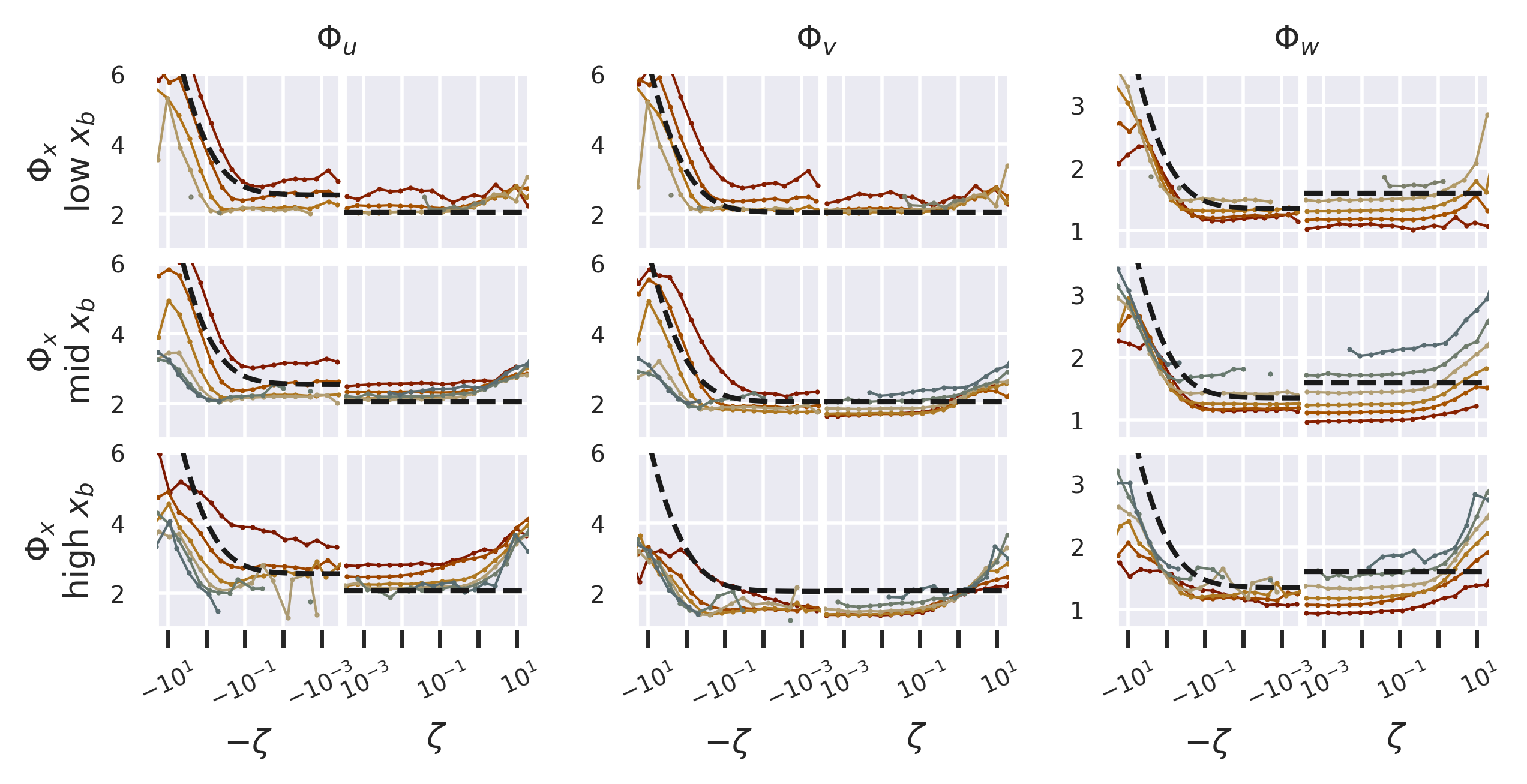}
\caption{Observed scaling for the square root of the variance of non-dimensionalized streamwise velocity $\Phi_u$ (left, a), spanwise velocity $\Phi_v$ (middle, b) and vertical velocity $\Phi_w$ (right, c) for different ranges of $x_b$: low $x_b$ ($0-0.3$), mid $x_b$ ($0.3-0.6$), and high $x_b$ ($0.6-1$). Lines are colored according to $y_b$ and show the median values across bins of anisotropy and $\zeta$.}
\label{fig:xblines}
\end{figure}

The correlation of the error with values of $x_b$ appears to be the result of consistent changes to the scaling and not just larger scatter. This becomes more obvious when examining the form of the scaling relations under different $x_b$ regimes as shown in figure \ref{fig:xblines}. For the scaling of streamwise and spanwise velocity, and to a lesser extent vertical velocity, we see similar behavior where for low values of $x_b$ (two component turbulence anisotropy) the different $y_b$ lines are scattered around the traditional MOST based scaling curve. As $x_b$ increases (becomes one component) the scaling based on $\zeta$ weakens while the $y_b$ scaling remains pronounced; turbulence under one-component anisotropy fails to scale significantly with $\zeta$. Under stable conditions for the streamwise and spanwise velocity, the scaling based on $\zeta$ is, by contrast, strongest under near one component regimes. It is worth noting, however, that the shapes of the curves for large $x_b$ values match fairly closely with those that one would expect based on the self-correlation between $\zeta$ and $\Phi_x$; in other words, closer to the type of curve expected purely based on the fact that $u_*$ is found on both axes. From examining the stable scaling we can also see that the scaling by $y_b$ is not consistent across values of $x_b$, particularly for streamwise velocity, which also helps explain poor performance of SC23 under stable conditions for this scaling. The findings in figure \ref{fig:xblines} appear consistent across sites, regardless of configuration or land cover.

\subsubsection{Drivers of Anisotropy}

\begin{figure}
\centering
\includegraphics[width=4in]{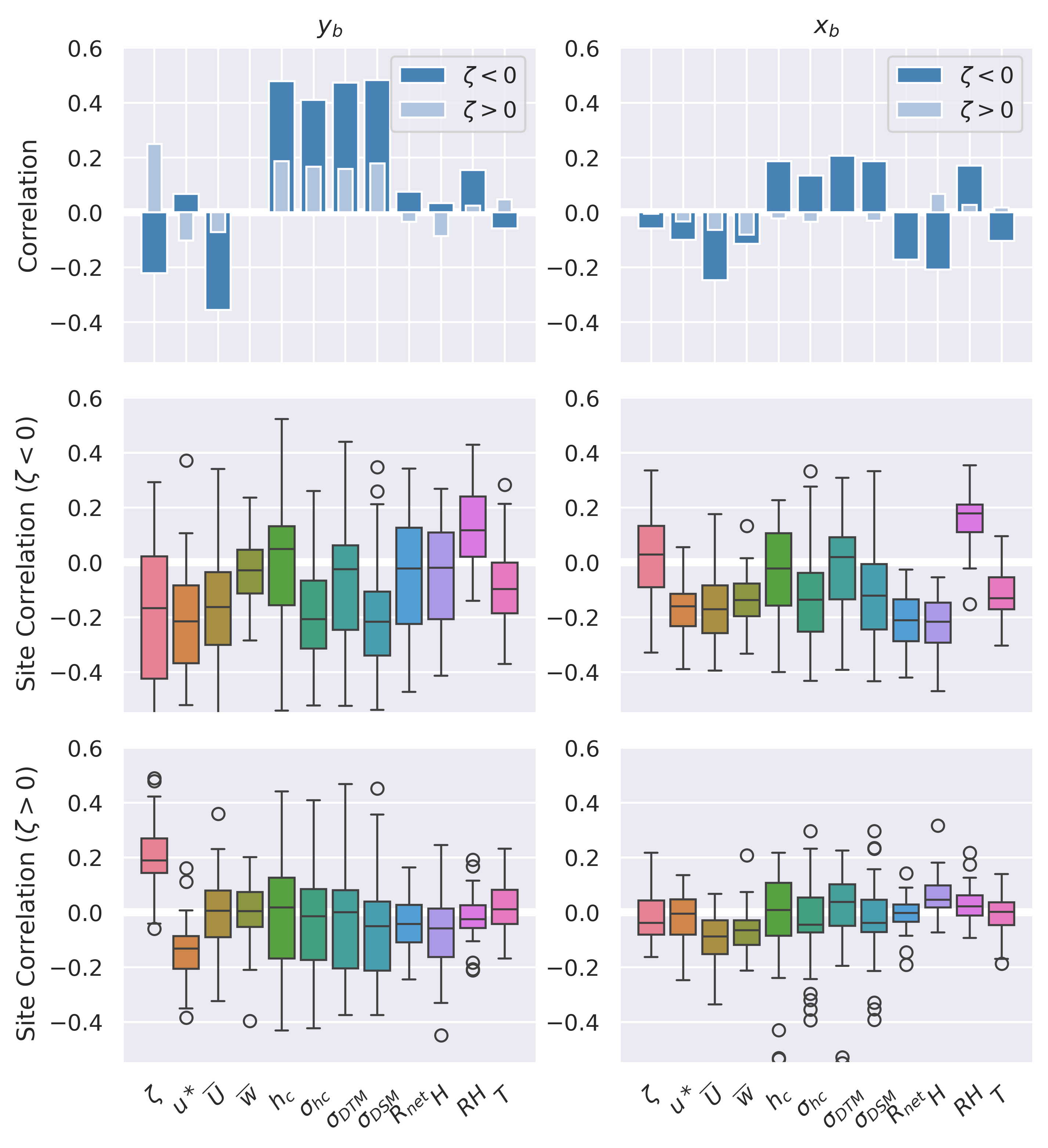}
\caption{Spearmann rank correlation of various variables with anisotropy $y_b$ (left) and $x_b$ (right). Variables include, left to right, stability $\zeta$, friction velocity $u^*$, mean horizontal wind speed $\overline{U}$, mean vertical velocity $\overline{w}$, median canopy height in flux footprint $h_c$, standard deviation of the digital terrain model in flux footprint $\sigma_{DTM}$, standard deviation of canopy height in flux footprint $\sigma_{h_c}$, standard deviation of the digital surface model in flux footprint $\sigma_{DSM}$, net radiation $R_{net}$, sensible heat flux $H$, relative humidity $RH$, and air temperature $T$. First row: correlation for unstable (dark blue) and stable (light blue) conditions across all sites. Second Row: boxplots of correlation coefficients at each individual site for unstable conditions. Third row: boxplots of correlation coefficients at each individual site for stable conditions.}
\label{fig:xbybcorr}
\end{figure}

This newfound relevance of $x_b$ requires a further exploration of the drivers of more one-component versus two-component turbulence anisotropy across NEON, as well as a comparison with potential drivers of the degree of anisotropy $y_b$. We begin a basic exploration in figure \ref{fig:xbybcorr} by examining the correlation between $x_b$/$y_b$ and various environmental variables for unstable and stable conditions. We examine the correlation for the 47 site ensemble (top panel) as well the correlation at each individual site (bottom two panels). Generally, correlations appear to be rather weak and inconsistent, revealing no obvious "smoking gun" relationship between these environmental variables and anisotropy. Nonetheless, some trends can illuminate our understanding of anisotropy and suggest paths for more in depth exploration. For both unstable and stable conditions, both in the entire dataset and when we examine it at each site, the variable most correlated to $x_b$ is $\overline{U}$. The correlation matches with previous studies reporting that one-component states are more common under lower wind speeds \citep{vercauteren_scale_2019,gucci_sources_2023}. 

Continuing to examine the degree of anisotropy, $y_b$, we see a fairly strong correlation between $y_b$ and four different metrics of complexity and vegetation: median canopy height $h_c$, standard deviation of the digital terrain model $\sigma_{DTM}$, standard deviation of the canopy height $\sigma_{h_c}$ and the standard deviation of the digital surface model $\sigma_{DSM}$ within the flux footprint, which serves as a metric for the combined complexity of the topography and canopy. This correlation is largely only present under unstable conditions when we examine the dataset as an ensemble of all sites. When we look at the correlation between $y_b$ and complexity at each individual NEON site, however, we get a wide range of correlation coefficients as evidenced by the bottom panels of figure \ref{fig:xbybcorr} as well as a median correlation less than zero, suggesting that within each site when the flux footprint includes more rough terrain and canopy, the value of $y_b$ decreases in direct contrast to the top panel of figure \ref{fig:xbybcorr} which implies the opposite. It is possible that the correlation we see in the top panel of figure \ref{fig:xbybcorr} is primarily driven by the fact that more isotropic turbulence (higher $y_b$) is more common over canopies and other rough surfaces (see figure \ref{fig:diurnal}). The high spread in correlation values at each individual site may also indicate that the metrics chosen for complexity are unable to account for larger scale topography, such as the location of the tower on a slope, ridge, or valley, which may impact the flow. It is notable that only the metrics that describe canopy complexity($\sigma_{DSM}$ and $\sigma_{hc}$) show consistent (albeit weak) inverse correlation with $y_b$, and topographic roughness ($\sigma_{DTM}$) does not appear to have this relationship. Ultimately, the inconsistency of the behavior indicates that a more thorough evaluation of the terrain and canopy at each site is necessary to fully evaluate the dependence of $y_b$ on complexity. The relationship between $y_b$ and surface/canopy complexity does not hold for stable stratification where no correlation is observed.

\subsubsection{Anisotropy across Temporal Cycles}
\begin{figure}
\centering
\includegraphics[width=3in]{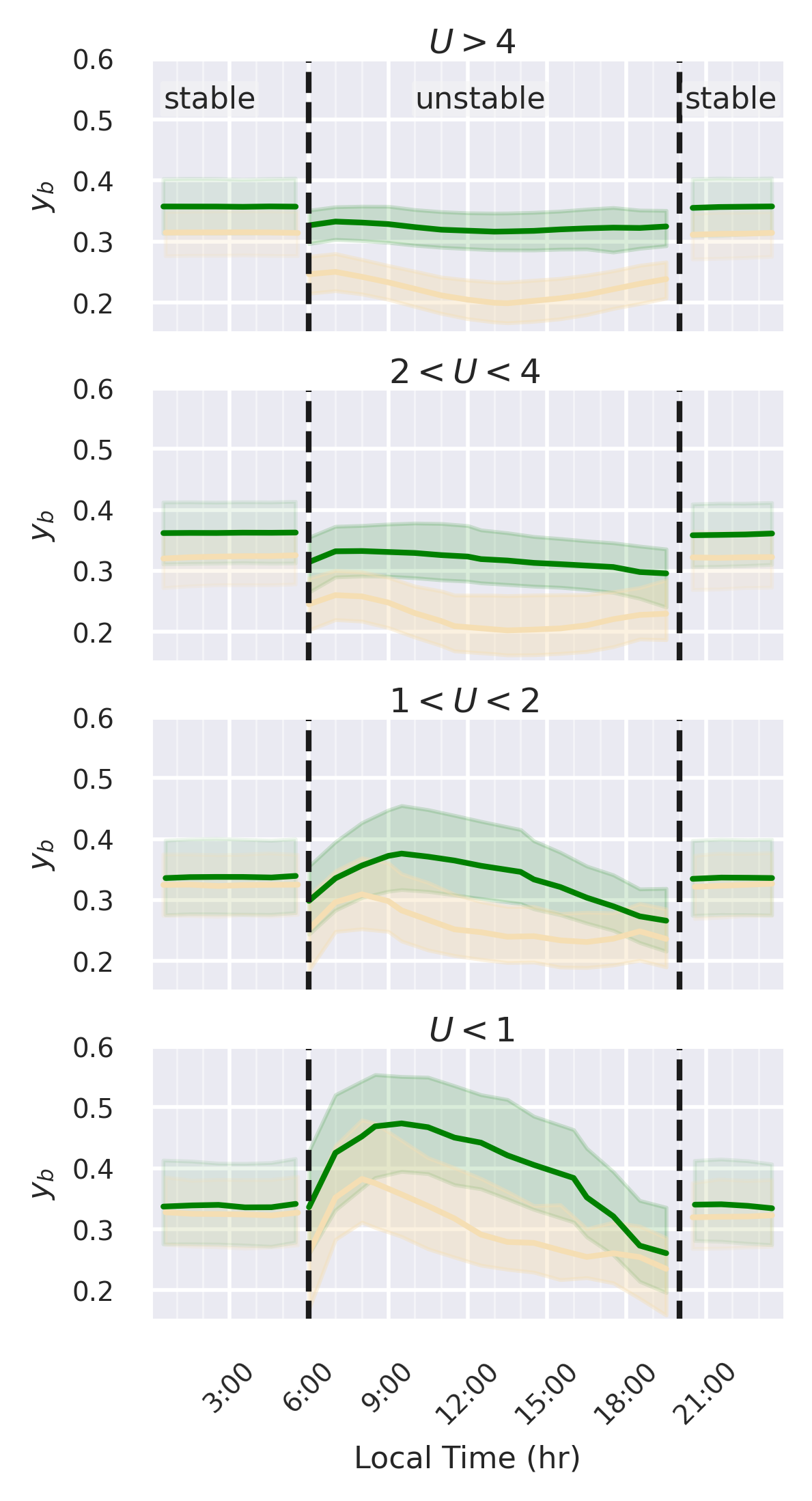}
\caption{Diurnal cycle of $y_b$ for forested sites (green) and non-forested sites (orange) with the median shown in full line, and the area between the 1st and 3rd quartile in shading. Plots show diurnal cycle for decreasing windspeed from top to bottom, starting with $\geq4\ m\ s^{-1}$, then between $2$ and $4\ m\ s^{-1}$, between $1$ and $2\ m\ s^{-1}$, and finally less than $1\ m\ s^{-1}$. Only unstable points are evaluated from 6:00 to 20:00 local time, and only stable points outside this period. Black lines separate the two regimes.} 
\label{fig:diurnal}
\end{figure}

One of the few variables with some consistent hints at correlation in figure \ref{fig:xbybcorr} is the mean wind velocity. The diurnal cycle of the degree of anisotropy under unstable stratification, presented in figure \ref{fig:diurnal}, illustrates some compelling differences in $y_b$ across wind speed and land cover (forested versus non-forested landscapes) throughout the day. First examining unstable stratification: For the lower wind speed regimes, there is a consistent diurnal cycle of more anisotropic turbulence in the very early morning, a peak in isotropy sometime during early to mid morning, a slow decay into the evening, and a collapse close to sunset. There is a clear separation of the sites with forest canopies and those without, with a consistent difference of $0.1 - 0.15$ in median daytime degree of anisotropy between them. For low wind speeds, there is also an offset in the timing of the most isotropic peak in the turbulence, occurring earlier (around 8:00 am) for the non-forested sites than the forested ones (between 9:00 and 10:00 am). This diurnal cycle weakens with higher wind speeds until it disappears almost entirely for forested sites under significant velocities, where the median approaches a near-constant value that is lower than the daytime average under lower wind conditions. For non-forested sites, however, a shifted diurnal cycle remains, with $y_b$ decreasing throughout the morning and increasing again during the evening. Notably, the variability around the median degrees of anisotropy also appears to tighten as velocities increase; this suggests that given some information about the site geometry (i.e. canopy), under high wind speeds and associated high shear conditions, one can reasonably estimate $y_b$. The physical drivers of this change are unclear, although one possible driver is changes in the depth of the atmospheric boundary layer, and therefore changes to critical flow scales which contribute to the anisotropy of turbulence (i.e. as vertical flow scales increase, turbulence may become more isotropic). This explanation, however, fails to explain the inverse diurnal cycle for non-forested environments under high velocities. A more focused investigation, including larger scale information not available at these sites, is likely required.
\par For stable stratification there is no clear diurnal cycle and only a weak separation according to land cover, with median degree of anisotropy varying between 0.3 and 0.375 with significant scatter. Despite this, it does appear that forested sites show higher $y_b$ values than non-forested sites, especially when velocities are higher. The low wind-speed analysis is somewhat complicated by the fact that theories of both models and measurements tend to fail under lower velocities when turbulence is less likely to be well developed. This can manifest in increased errors and deviations from scaling curves under low velocity conditions. Examining this directly yields expected results; in supplementary figure S5 we see both MOST and (to a lesser extent) anisotropy modified scaling show worse performance at low velocities. In general, and especially under unstable conditions, we see MOST performance decay more quickly with wind speed whereas the anisotropy generalized relations are able to maintain decent performance.

\begin{figure}
\centering
\includegraphics[width=4.5in]{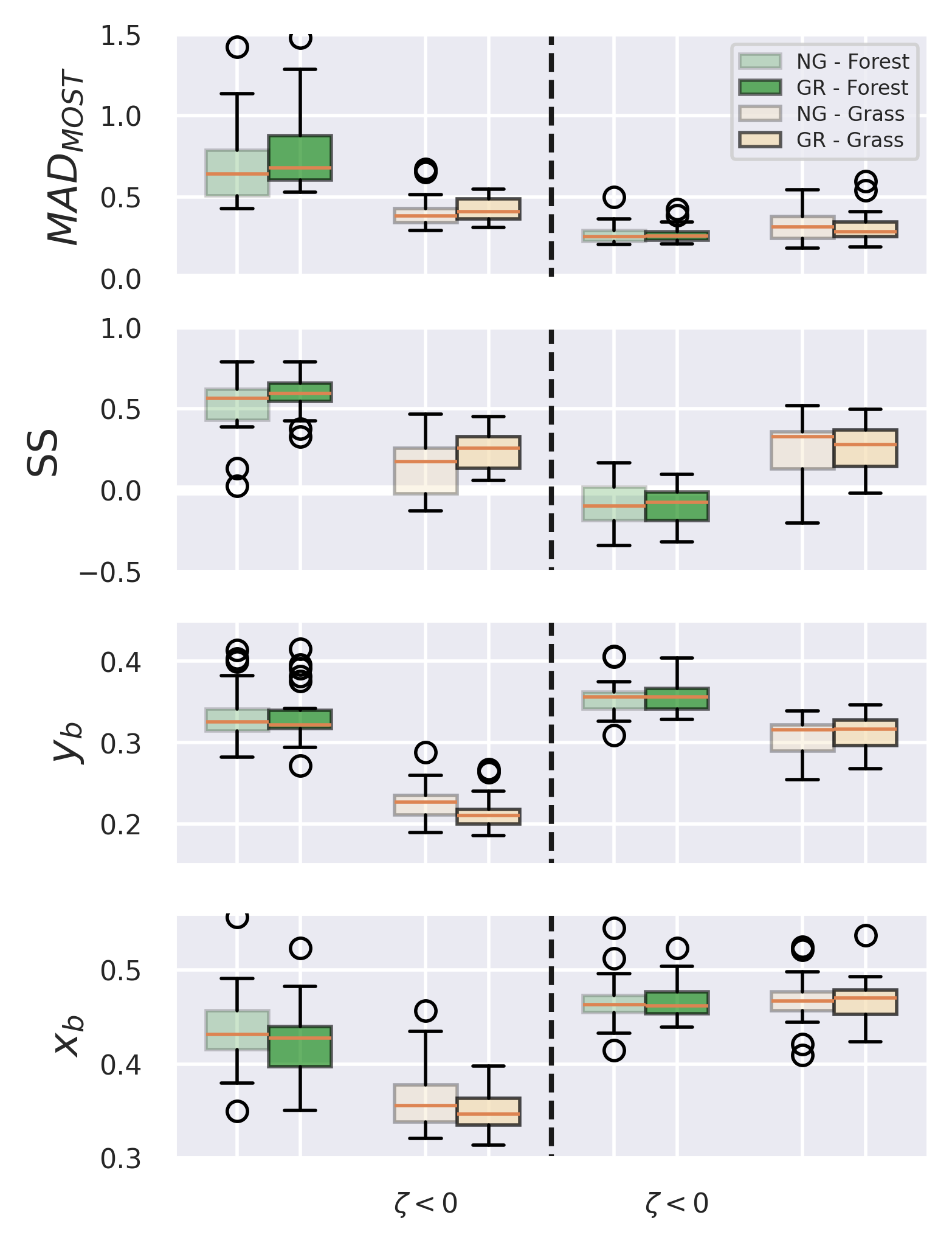}
\caption{Box-whisker plots showing the median values across the 47 sites for, from top to bottom, the median absolute deviation of MOST from the data, the skill score of the refit when compared to traditional most, $y_b$ and $x_b$. Boxplots are shown for unstable conditions (left half) and stable conditions (right half), for forested terrain during the growing season (GR; dark green) and outside the growing season (NG; light green), as well as for non-forested sites during (GR; tan) and outside (NG; light tan) the growing season.}
\label{fig:growing}
\end{figure}

The trend of higher $y_b$ at forested sites when compared to non-forested sites appears to be consistent seasonally as well. Figure \ref{fig:growing} shows how the $MAD$ for $\Phi_u$ from the traditional MOST relations, the refit skill score, $y_b$ values and $x_b$ values change between the growing season and outside the growing season for forested and non-forested sites. As in figure \ref{fig:diurnal}, we see consistent differences in $y_b$ between the two broad ecosystem categories. We also see differences in $x_b$ between forest and grassland terrain, however this only appears to apply for unstable conditions. When we examine these trends across seasons (during and outside of the growing season), we see some small but significant trends. Under unstable conditions, $y_b$ and $x_b$ appear to trend slightly lower during the growing period as compared to outside the growing period. It may be that leaves generally lead to more aerodynamically homogeneous canopies as opposed to isolated branches or low lying vegetation, generally causing $y_b$ to be more reflective of the flatter homogeneous sites. There are also seasonal changes in performance; at both grassland and forest sites the error of traditional MOST increases during the growing season. This increased error, however, is paired with an increased skill score for the anisotropy modified relations, suggesting that the $y_b$ is able to absorb the influence of vegetation changes on surface layer scaling. Indeed, we see $MAD_{MOST}$ increase by approximately 15\% while $MAD_{y_b}$ increases by less than 1\%. The same trends exist, only magnified, when isolating deciduous forests (not shown) where the seasonal changes in vegetation are more pronounced. 

\subsection{Inter-site Variability}
\begin{figure}
\centering
\includegraphics[width=5.5in]{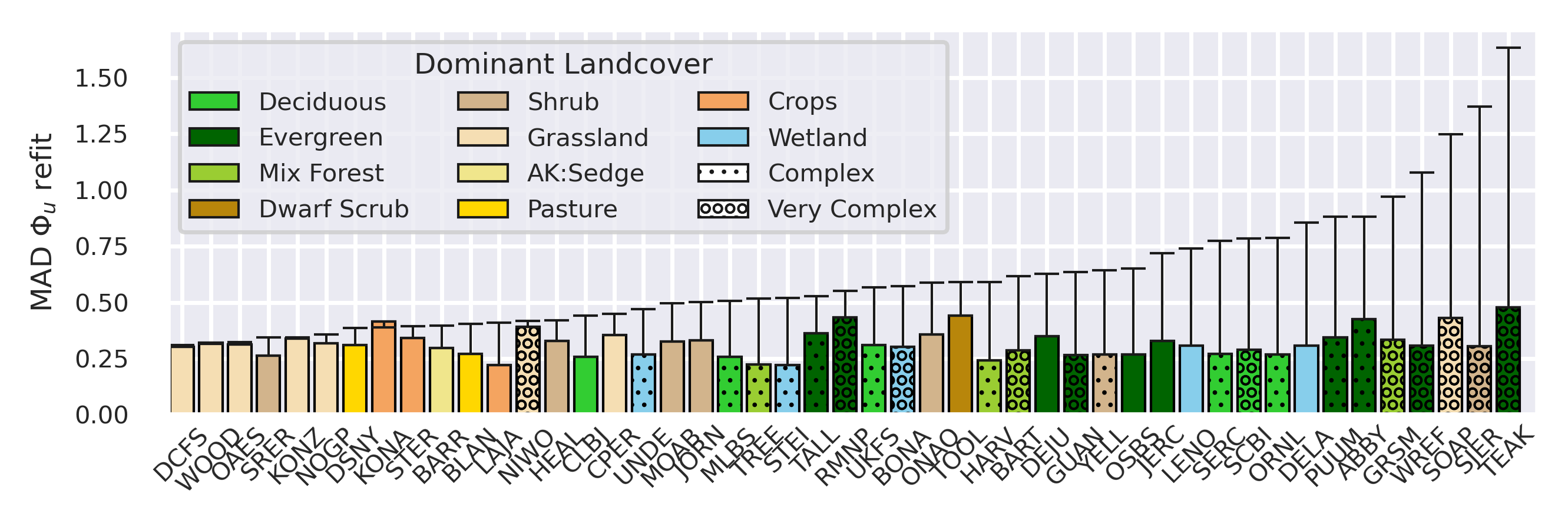}
\caption{Median Absolute Deviation (MAD) for $\Phi_u$ under unstable conditions for the anisotropy modified relations (colored bar) and for traditional MOST (black tick lines) for each site in NEON. Sites are colored according to dominant landcover and organized from lowest traditional MOST error (MAD) to highest. Sites with high complexity (median standard deviation of the digital surface model (DSM) in the flux tower footprint$>10$) are marked with black dots, and very high complexity (median standard deviation of the digital surface model (DSM) in the flux tower footprint$>20$) marked with black circles.}
\label{fig:sitebar}
\end{figure}

Given the large surface variability present across NEON, including wetlands, grasslands, deserts, sparse alpine forests, arctic coastlines and tropical island hillslopes, it is promising to see the relatively consistent performance of the SC23 relations. There are, however, some differences across the sites worth exploring. In figure \ref{fig:growing}, there are consistent differences under stable and unstable stratification between the performance of anisotropy generalized relations, with grasslands showing slightly larger skill scores under stable conditions and lower skill scores under unstable stratification. We can explore this further; figure \ref{fig:sitebar} shows the performance of both MOST and SC23 for the scaling of streamwise velocity under unstable conditions across the network. It should come as no surprise that the performance of traditional MOST is high for relatively flat grasslands, crop and pasturelands and that SC23 provides only marginal improvements for these sites; those sites are largely representative of the locations where MOST was originally developed and hew more closely to MOST assumptions of planar homogeneity. Their climatology is likely also more driven by shear conditions (two component anisotropy) and the tower heights tend to be lower, which is more representative of classic MOST \citep{stiperski_scaling_2019}. Grassland, pasture or crop sites that do see more significant improvements from SC23 (BLAN, LAJA) have other complicating factors. BLAN and LAJA both have small forest stands present in their flux tower footprint depending on the direction of the mean flow, which adds flow complexity that SC23 seems to account for while traditional MOST appears to fail.

The performance improvements from generalizing MOST with anisotropy are largely concentrated in the most complex and forested sites, which are also the sites with the highest deviation from traditional MOST and deviate in typical patterns of anisotropy from the flat, non-forested sites (see figure \ref{fig:diurnal}). To represent complexity both in canopy coverage and topography, we use the standard deviation of the digital surface model (DSM) within the flux tower footprint; see table \ref{tab:sites} for the complexity categories. The DSM shows the elevation from lidar, and includes both topography and canopy. While an imperfect metric of complexity, it can highlight sites that may appear simple based on solely the land cover type in figure \ref{fig:sitebar}. The reduction in MAD by accounting for anisotropy in MOST are most significant at some of the most complex sites; SOAP and SJER for example, are both sites with very sparse canopy over complex topography. GRSM and ABBY features a mix of short and tall canopy with deciduous and evergreen trees at complex topography. WREF and TEAK are both very tall canopies in mountainous regions. The combined effect of little changes over flat homogeneous terrain and stronger improvements in more complex forested terrain is that the SC23 relations produce end results of relatively consistent performance across nearly all ecosystem types. The anisotropy modified relations bring all sites $MAD$ below $0.5$ whereas only two forested sites have $MAD$ this low using traditional relations. This illustrates that the relations are not only robust across varied terrain but do not appear to show any strong consistent focus towards any particular type of ecosystem - a desirable trait for future implementation in regional and global models tasked with representing surface exchange equally in flat grasslands and complex forests. 

\begin{figure}
\centering
\includegraphics[width=5.5in]{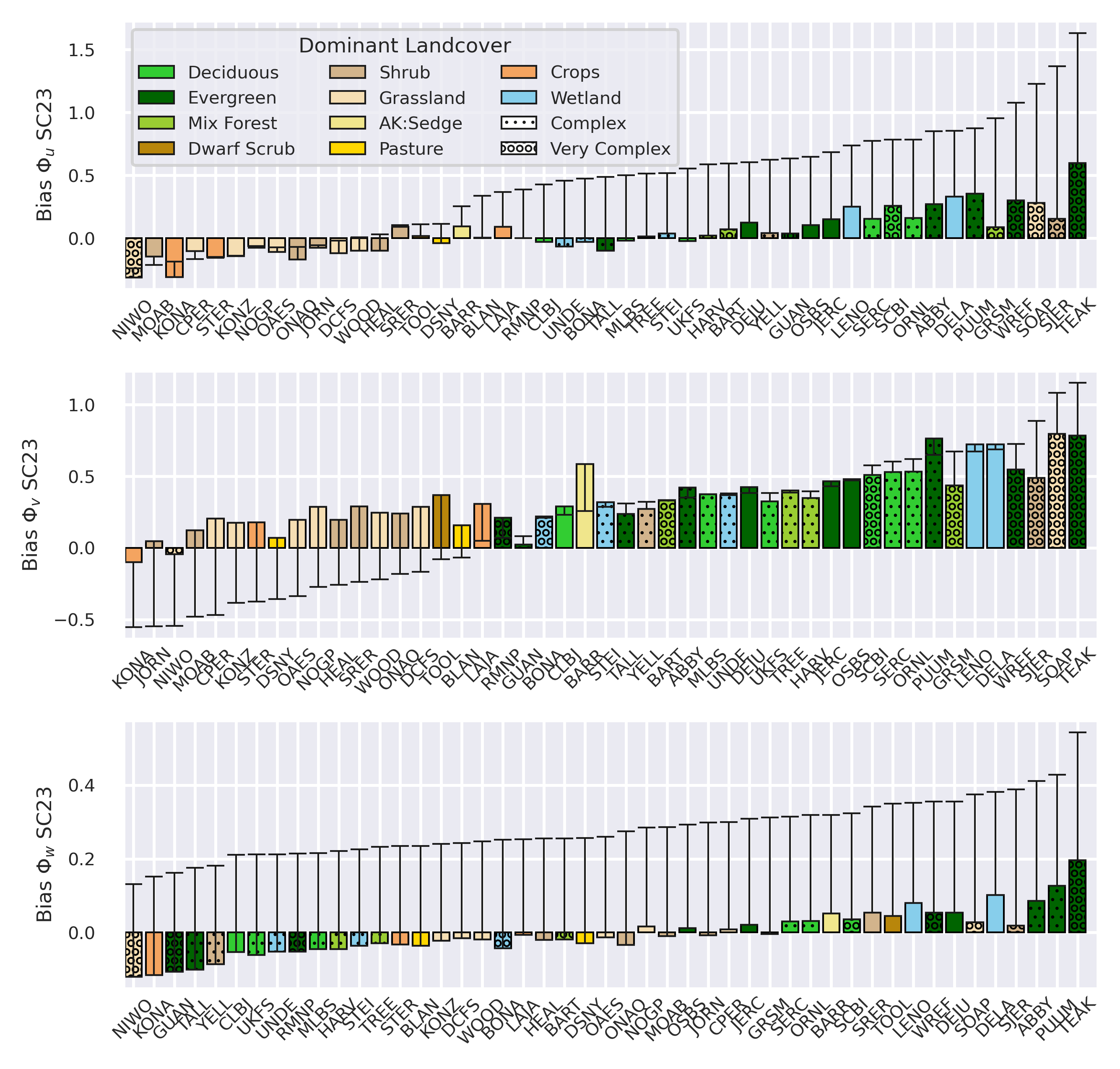}
\caption{Median bias for $\Phi_u$ (top), $\Phi_v$ (middle), and $\Phi_w$ (bottom) under unstable conditions for the SC23 relations (colored bar) and for traditional MOST (black tick lines) for each site in NEON. Sites are colored according to dominant landcover and organized from lowest traditional MOST bias to highest. Sites with high complexity (median standard deviation of the digital surface model (DSM) in the flux tower footprint$>10$) are marked with black dots, and very high complexity (median standard deviation of the digital surface model (DSM) in the flux tower footprint$>20$) marked with black circles.}
\label{fig:sitebarbias}
\end{figure}

$MAD$ provides a clear illustration of error, but that error can originate from a biased scaling relation or simply scatter in the data. As such, we now examine the bias of the $\Phi_u$ scaling across the NEON sites. The first panel of figure \ref{fig:sitebarbias} shows that there are a few consistent differences between the forested and non-forested sites at least in the bias of the generalized and traditional MOST relations. At about half of the sites with forest cover, the absolute bias is reduced using generalized MOST over classic MOST to the level of the homogeneous, flat, non-vegetated sites but the bias at forested sites largely remains positive, whereas the bias at many of the flat homogeneous sites remains negative. The forested sites overall see a much greater bias correction by accounting for $y_b$ when compared to the flat homogeneous sites where the bias tends to remain the same or has slight increases/decreases. 

Bias in the scaling of $\Phi_v$ shows similar patterns to $\Phi_u$ across the sites, with a few small differences. Key is that the SC23 relations only improve bias over non-vegetated sites in the network, which helps explain why the SC23 performance over NEON did not match that over the original (non-vegetated) sites analyzed in \citep{stiperski_generalizing_2023}; this trend is also observed under stable conditions. SC23 scaling of $\Phi_v$ under unstable conditions is not extendable to complex, forested terrain without refitting the parameters unlike the $\Phi_u$ and $\Phi_w$ scaling which are immediately extendable. The refit parameters from this work (presented in table S1), successfully correct for this bias. The scaling for $\Phi_w$ under unstable conditions does not display the same grouping (with non-vegetated and vegetated sites grouped together) and it appears instead that bias is lowest for complex, non-vegetated (or with short canopies) sites and highest for complex, high and/or sparse canopies. SC23 corrects a persistent bias in traditional MOST, but largely retains this trend of lower (negative) bias for complex, non-vegetated and higher (positive) bias for tall or sparse canopies. We can also examine the bias of the stable scaling relations across the NEON sites; these are shown in figures S6. Bias patterns for stable scaling relations are largely similar to those seen in the unstable scaling, with the degree of improvement adjusted as would be expected from figure \ref{fig:SS_all}. The SC23 curves show a persistent negative bias for $\Phi_u$ under stable conditions that explains the poor performance under these conditions. 

\section{Discussion}
\label{sec:discussion}
\subsection{Relationship Between Canopy, Performance of Scaling Relations, and Anisotropy}
While previous work has proven that anisotropy can be used to generalize MOST to cover complex topographies, with this work we can now show that this generalization extends easily to vegetated canopies, ecosystems inside and outside the growing period, lower wind conditions than MOST, and in complex non-homogeneous ecosystems. For the scaling of $\Phi_u$ and $\Phi_w$, the performance of the SC23 relations over the NEON tower network, which contains mostly vegetated canopies, matched the results over the initial non-forested network over which the results were fit, showing that they were immediately extendable to these conditions where more isotropic turbulence is more prevalent. Particularly under unstable conditions, the SC23 relations were able to correct for a persistent bias over more complex terrain. The scaling of $\Phi_v$ from SC23, however, required refit parameters to extend performance improvements to cover canopies as the original SC23 fits only improved performance on non-vegetated sites. Refitting to the NEON dataset, in the case of $\Phi_v$ scaling, was shown to bring performance improvements back to the level observed in previous studies \citep{stiperski_generalizing_2023}, as $\Phi_v$ was shown to be lower over canopies, and to a lesser extent non-vegetated terrain, in our work. The cause of this difference in the scaling of $\Phi_v$ remains unclear. 

The anisotropy, mostly through $y_b$, appears to be modulated by canopy and terrain complexity for unstable conditions, but the relationship remains unclear and likely depends on details of the configuration at each individual site as evidenced by the wide spread in figure \ref{fig:xbybcorr} site correlations, as well as their low overall value. What these results do communicate is that, contrary to previous hypotheses, the variability of the canopy and topography within the flux footprint of the tower at each site does not directly translate to differences in anisotropy at most sites, however the mean site complexity and canopy coverage does influence anisotropy (see the difference between the overall correlation in the top of figure \ref{fig:xbybcorr} when compared to the site correlation in the bottom two panels). The comparison between leaf on and leaf off periods in figure \ref{fig:growing} further supports the idea that $y_b$ may decrease with increased vegetation on a local scale. This apparent discrepancy may be caused by a number of overlapping factors. The metrics we use in this study for topographic and canopy complexity are rather simplistic, and may be missing out on other vital characteristics of site geometry as they relate to the flow; is the site on a peak or in a valley? What is the orientation of the wind relative to the slope? What is the aspect of the slope relative to solar azimuth? What is the vertical canopy structure and how is it influencing the profile of wind velocity? These are all questions that the simplistic metrics do not answer, and a more in depth analysis of site geometry would be needed to account for these effects. In addition, the tower footprint scale may not be the primary geometric scale impacting anisotropy, which could be influenced more by larger scale topographic structures. It is also a challenge to separate potential impacts of (possibly) being inside of the roughness sub-layer, where fundamental MOST scaling assumptions do not apply. It remains unclear if the anisotropy varies throughout the roughness sublayer in such a way that anisotropy generalized MOST scaling can effectively counter these deviations from core assumptions or not, and although the NEON network was designed to have the 3D sonic anaemometer lie above the roughness sublayer, it cannot be guaranteed without more focused analysis as NEON does not contain a vertical profile to fully evaluate changes in the scaling with height.

While more needs to be done to fully understand the impact of canopy and topography on anisotropy, this work shows that there is a clear difference between vegetated and non-vegetated sites that appears to be both captured by anisotropy generalization of MOST, and has potential to be impactful in model parameterizations. An outstanding problem that remains for application of SC23 and the adjustments presented in this work to large scale NWP models and ESMs is the fact that $y_b$ (and, if utilized, $x_b$) need to be determined a-priori, or iterated upon. At present, there is no established method to compute $y_b$ in the surface layer for large scale models. The strong and consistent separation in $y_b$ for forested versus non-vegetated sites, as well as the consistent diurnal cycle modulated by wind velocity (figure \ref{fig:diurnal}), provides potential for a first order estimation of $y_b$ based on only wind speed and land cover type, both readily available parameters in ESMs. This also suggests avenues of future work to build upon to refine the observed diurnal variability and develop an improved parameterization of anisotropy. 

\subsection{Scaling of One-Component versus Two-Component Anisotropy}
Over NEON we were able to identify a complex, nuanced relationship between $x_b$ and the observed scaling. This examination of $x_b$ also provides a deeper insight into the fundamental behavior of turbulence in the surface layer. Under unstable stratification, we see consistently weaker scaling for more one component anisotropy, especially in the free convective regime. For stable conditions, we see a different trend with scaling with $\zeta$ appearing strongest under one component anisotropy. This increase begins to rise towards the curve that would be expected due to the self-correlation of $\Phi_{u_i}$ with $\zeta$, which yields another physical interpretation: that under predominantly one-component anisotropy, self-correlation appears strong and $\Phi_u$ and $\Phi_v$ do not scale physically with degree of stratification. Under predominantly two component anisotropy, the observed scaling with stratification appears weaker (indicative of z-less scaling for the majority of the stability range), but is further from the curves expected due to self-correlation (the 1/3rd power law) implying greater physical scaling. This implies that $\zeta$ based scaling, whether MOST or SC23, start to fail under highly one component turbulence anisotropy. The results also hint at the potential for an $x_b$ based scaling. This is particularly true for $\Phi_u$ and $\Phi_v$ under stable conditions, where there is a stronger error dependence on $x_b$ than $y_b$, and a clear, simple, observable pattern of behavior. For the other scaled variables, $y_b$ still appears to play the dominant role, however a scaling based on both $y_b$ and $x_b$ has the potential for improvements, particularly in circumstances where one-component turbulence anisotropy is expected to be prevalent.

\subsection{Self Correlation}
A question that remains with the SC23 relations, and the slight modifications proposed as part of this work, is the relative impact of self correlation on the apparent relations. As shown in equation \eqref{yb}, $y_b$ is a function of $\overline{u_i’u_i’}$, and $\Phi_{u_i}$ is also, of course, a function of $\overline{u_i’u_i’}$. This issue of self correlation has an impact on all MOST based scaling making the SC23 relations certainly not unique in this respect \citep{l_andreas_comments_2002,baas_exploring_2006}; $\Phi_{u_i}$ and $\zeta$ are both functions of $u_*$ for example. Observed relations may be due to self-correlation, or a consequence of the physical scaling of the variances and the defined curves. Previous work provides a method for evaluating degree of self-correlation by randomizing each independent variable in the relation so that all physical connection between the variables is lost. Any observable relation using the randomized data is almost certainly the result of self-correlation, which can then be compared to the relation observed with the real data to evaluate the degree to which self-correlation is responsible for the observed anisotropy dependence of $\Phi$. We randomized all elements of the Reynolds stress tensor ($\overline{u'u'}$, $\overline{v'v'}$, $\overline{w'w'}$, $\overline{u'v'}$, $\overline{u'w'}$, $\overline{v'w'}$) and used these values to compute $\Phi_{u}$, $\Phi_{w}$ and $y_b$. The random and real scaling of $\Phi_{u_i}$ for unstable and unstable conditions is shown in figure S7. The randomized data for all relations has significantly more scatter than the real data, especially for the scaling of $\sigma_u$; even if we try to fit a curve to this highly scattered data the curve does not fully match up with the curves for the real data. Spearman correlation coefficients for the real data curves are consistently larger than the randomized data; for $\Phi_u$ under unstable stratification, $r_{real}=0.56$ compared to $r_{random}=-0.03$. Under stable conditions, $r_{real}=0.10$ compared to $r_{random}=-0.04$. For $\Phi_w$, self correlation is a factor, but physical scaling still plays a significant role with $r_{real}=0.46$ and $r_{random}=0.33$ under unstable conditions and $r_{real}=0.52$ and $r_{random}=0.12$ under stable stratification. In general, however, we see through this examination of the degree of self-correlation between $y_b$ and $\Phi_{u_i}$ that surface layer physics is driving the results from the anisotropy generalized MOST scaling relations.

\section{Conclusion}
\label{sec:conclusion}
An anisotropy based generalization of MOST shows consistent improvement in performance over traditional MOST scaling for complex sites without canopy, over which it was developed, as well as sites with canopy, as newly shown in this work. The SC23 framework, with only limited alteration for the scaling of $\Phi_v$, readily extends across sites ranging from arctic coastline, to flat deserts and from rolling oak savannas, to dense temperate canopies. We find a consistent difference in the typical anisotropy of turbulence between these sites, which the modified versions of the SC23 framework readily capture. This difference in behavior that could potentially be leveraged, along with differences based on wind speed, to parameterize anisotropy in large scale models, which currently lack sufficient information to compute anisotropy directly. We also show through this work the role that $x_b$ plays in modulating $\zeta$ based ASL scaling from both SC23 and MOST, particularly its potential to be the dominant anisotropy invariant (over $y_b$) for understanding the scaling of horizontal velocities under stable stratification. We saw consistent weakening of $\zeta$  based scaling when $x_b$ was high, further demonstrating the challenges of applying traditional similarity theory under highly one-component turbulence anisotropy. While additional efforts are necessary to further advance understanding of the anisotropy of turbulence near the surface and the interplay between ASL scaling, anisotropy and canopies, this work illustrates the potential of anisotropy based scaling and parameterizations to enhance existing MOST schemes, reduce scatter, and supplement existing surface layer parameterization schemes in atmospheric models.

\section*{acknowledgments}
The authors thank the various agencies that have supported this work, including the National Science Foundation Division of Atmospheric and Geospace Sciences (NSF-AGS) Postdoctoral Fellowship (Award 2412560; PI Dr. Tyler Waterman), NSF-AGS Physical and Dynamic Meteorology Program (PDM) (Award 2414424; PI Dr. Marc Calaf) and the European Research Council (ERC) (Award 101001691; PI Dr. Ivana Stiperski).

\printendnotes
\bibliography{mybib}

\end{document}


\maketitle

\section{Supplemental Figures and Tables}

\begin{table}[t]
\caption{Table showing the coefficient values for the polynomial fits of the anisotropy dependence for the $a(y_b)$ and $b(y_b)$ in table 1. Form of the polynomial is $\Sigma_{n=0}^3 p_n(y_b)^{n}$ or $\Sigma_{n=0}^3 p_n(\log_{10} y_b)^{n}$ when specified. The SC23 values for these parameters are found in the supplementary of SC23. Notably, we choose to scale $a$ in $\Phi_w$ under unstable stratification with $\log_{10} y_b$ in contrast to $y_b$ used in SC23, making the parameters for these incomparable.}
	\label{tab:params}
	\begin{center}
	\begin{tabular}{c c c c c c}
		\hline
		\begin{tabular}{@{}c@{}} $\Phi_x$ \\ \textbf{Unstable} \end{tabular} & Param. & $p_0$& $p_1$& $p_2$& $p_3$ \\ \hline
            $\Phi_u$ & $a(\log_{10} y_b)$ & 0.532 & -2.955 & --- & --- \\ \hline
            $\Phi_v$ & $a(\log_{10} y_b)$ & 1.072 & --- & 2.632 & --- \\ \hline
            $\Phi_w$ & $a(\log_{10} y_b)$ & 0.932 & -0.620 & -0.516 & ---  \\ \hline \hline
		\begin{tabular}{@{}c@{}} $\Phi_x$ \\ \textbf{Stable} \end{tabular} & Param & $p_0$& $p_1$& $p_2$& $p_3$ \\ \hline
            $\Phi_u$ & $a(y_b)$ & 3.590 & -6.912 & 8.129 & --- \\
             & $d(y_b)$ & 0.055 & --- & --- & --- \\ \hline
             
            $\Phi_v$ & $a(y_b)$ & 1.672 & -0.567 & 2.258 & --- \\
             & $d(y_b)$ & 0.193 & -0.244 & --- & --- \\ \hline
             
            $\Phi_w$ & $a(y_b)$ & 0.949 & 0.007 & 2.643 & --- \\
             & $d(y_b)$ & 0.077 & --- & --- & --- \\ \hline

        \end{tabular}
	\end{center}
\end{table}

\subsection{Determining Averaging Period}
\begin{figure}
\centering
\includegraphics[width=4in]{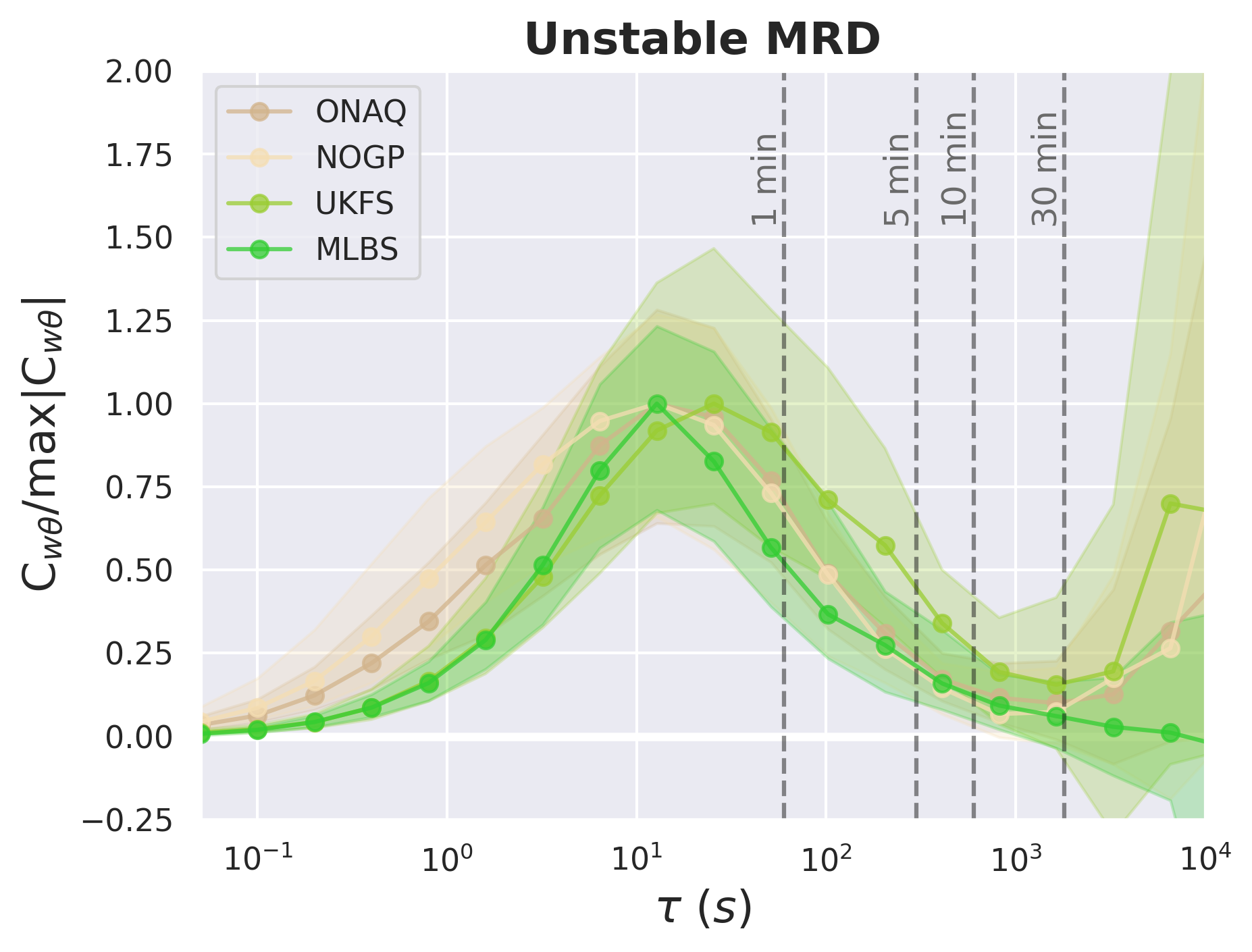}
\caption{Multi-resolution decomposition of the sensible heat flux under unstable conditions for four representative NEON sites, with interquartile range for the MRDs filled in. Time scale $\tau$ shown in log scale.}
\label{fig:mrd_unstable}
\end{figure}

Multi-resolution decomposition (MRD) is a useful tool for determining appropriate averaging periods as it can isolate the scale-wise contribution to the overall flux, and identify the spectral gaps (and, in the case of $w'\theta'$ the point where the lines cross zero) that are usually chosen for appropriate averaging periods and the differentiation between turbulent and non-turbulent motions. MRD is computationally expensive, so we only perform it for one year at 4 representative sites including complex and flat forest and non-vegetated terrain. Figure \ref{fig:mrd_unstable} shows that there is a spectral gap for all 4 sites between 10 and 30 minutes, reinforcing the choice of 30 minutes as an appropriate averaging period.

\begin{figure}
\centering
\includegraphics[width=4in]{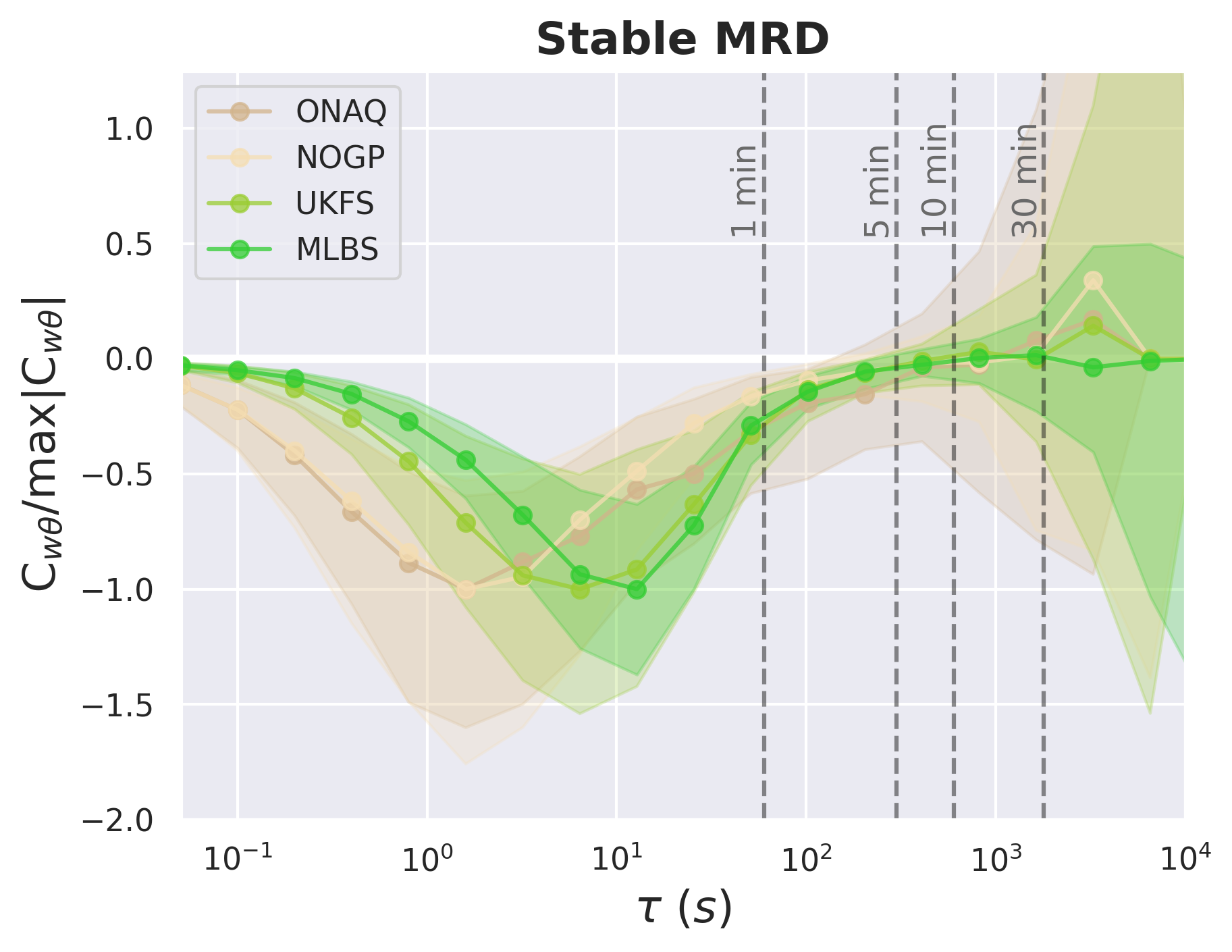}
\caption{Multi-resolution decomposition of the sensible heat flux under stable conditions for four representative NEON sites, with interquartile range for the MRDs filled in. Time scale $\tau$ shown in log scale.}
\label{fig:mrd_stable}
\end{figure}

Under stable conditions (figure \ref{fig:mrd_stable}), the spectral gap/zero crossover point lies somewhere between 5 and 15 minutes. In this case, however, there is significant likelihood of positive $w'\theta'$ contributions after 5 minutes, which are, by definition, not "stable". This poses issues for averaging periods longer than 5 minutes. Additionally, at 5 minutes, nearly all of the major flux contributions are already captured, meaning additional averaging would have marginal impact on capturing the turbulent flux with higher risk for erroneous contributions that would impact fluxes, variances, and anisotropy.

\subsection{Anisotropy and $\zeta$ Distribution}

\begin{figure}
\centering
\includegraphics[width=4in]{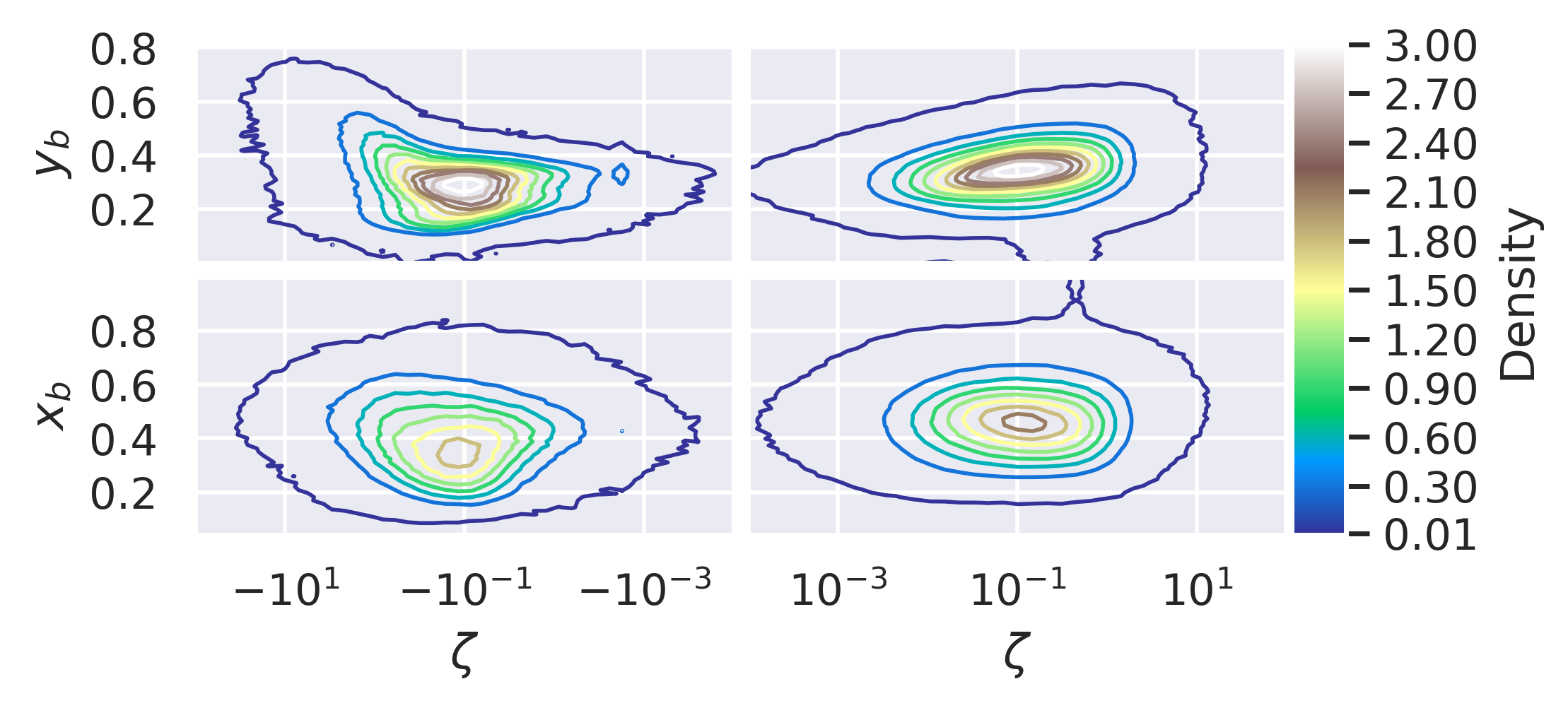}
\caption{2D histogram density contour comparing stability $\zeta$ to anisotropy invariants $y_b$ (top) and $x_b$ (bottom). Left side of figure shows unstable stratification and the right shows stable stratification. Density such that the area under the histogram integrates to 1.}
\label{fig:zL_ani}
\end{figure}

\subsection{Additional Analysis}
\begin{figure}
\centering
\includegraphics[width=4in]{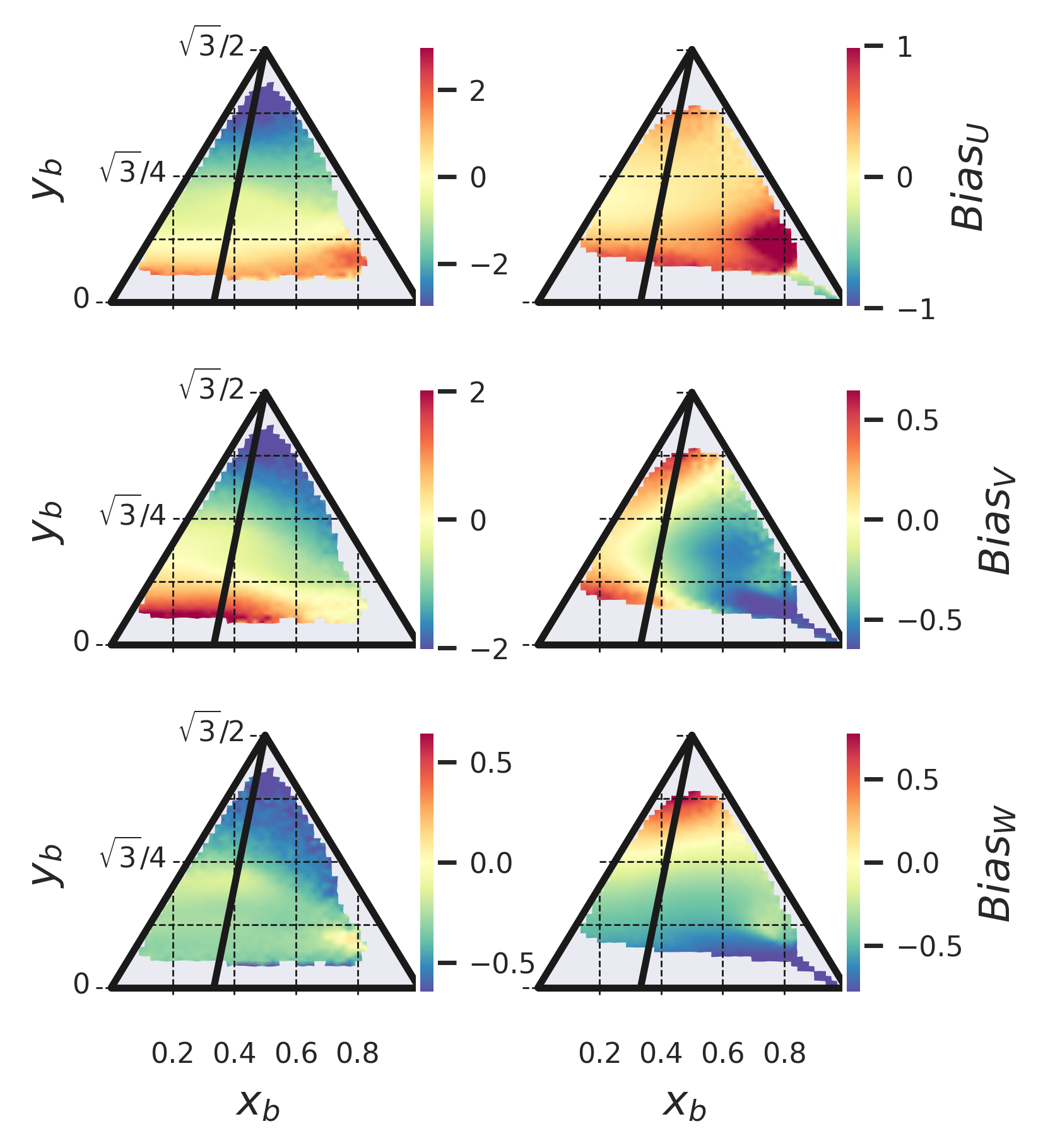}
\caption{Bias of traditional MOST relations to the data mapped onto the Barycentric Map of the Lumley Triangle. Blue signifies lower bias and red higher bias. Bias shown for unstable (left) and stable (right) conditions for the variances of u (top) v (middle) and w (bottom). The plane-strain line is plotted in black for reference.}
\label{fig:lumleybias}
\end{figure}

\begin{figure}
\centering
\includegraphics[width=4in]{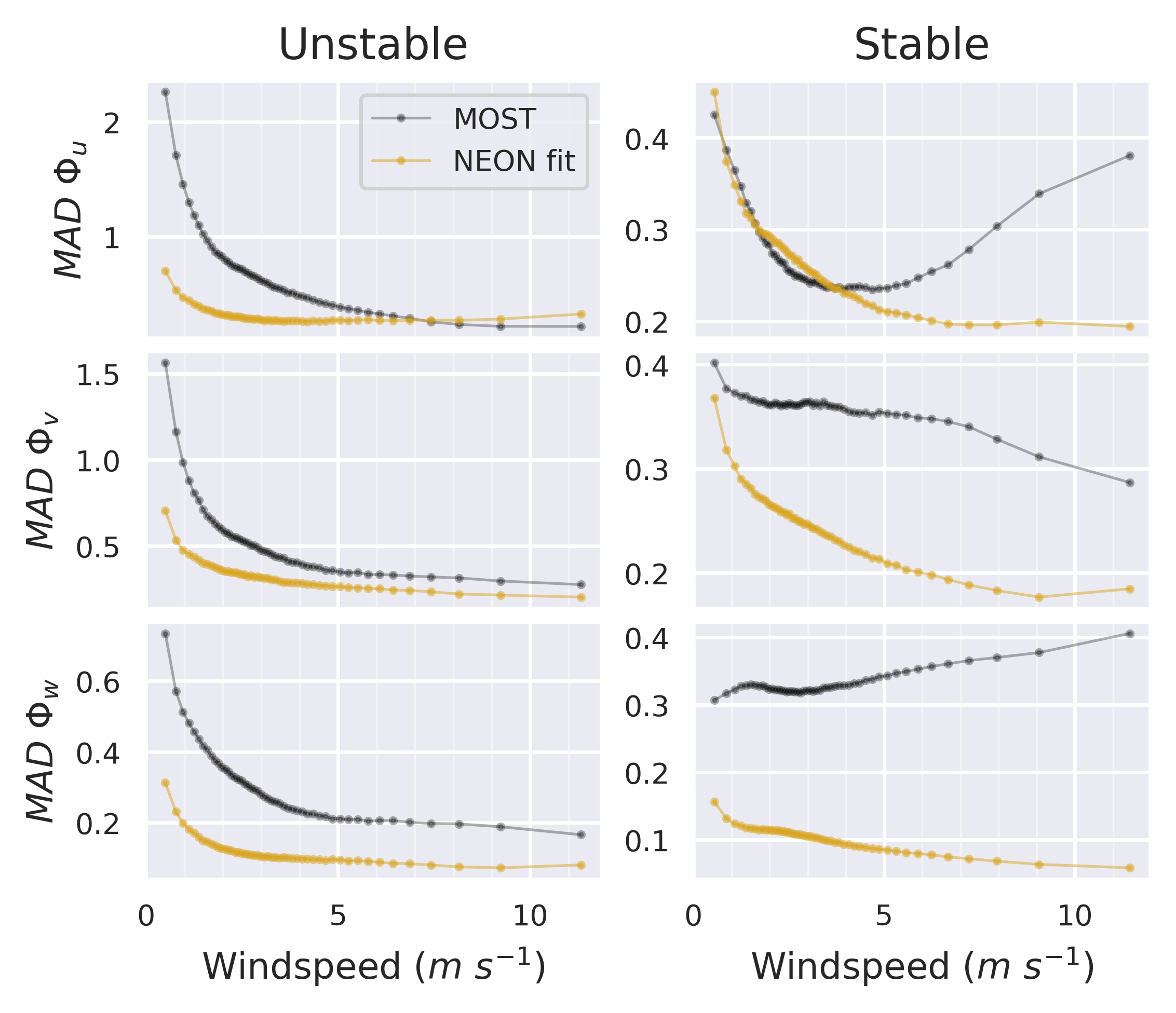}
\caption{$MAD$ of MOST relations (black) and the NEON refit (orange/yellow) for $\Phi_u$, $\Phi_v$ and $\Phi_w$ under unstable (left) and stable (right) stratification, binned into 50 equal sample-size bins based on mean wind speed.}
\label{fig:windmad}
\end{figure}

\begin{figure}
\centering
\includegraphics[width=5.5in]{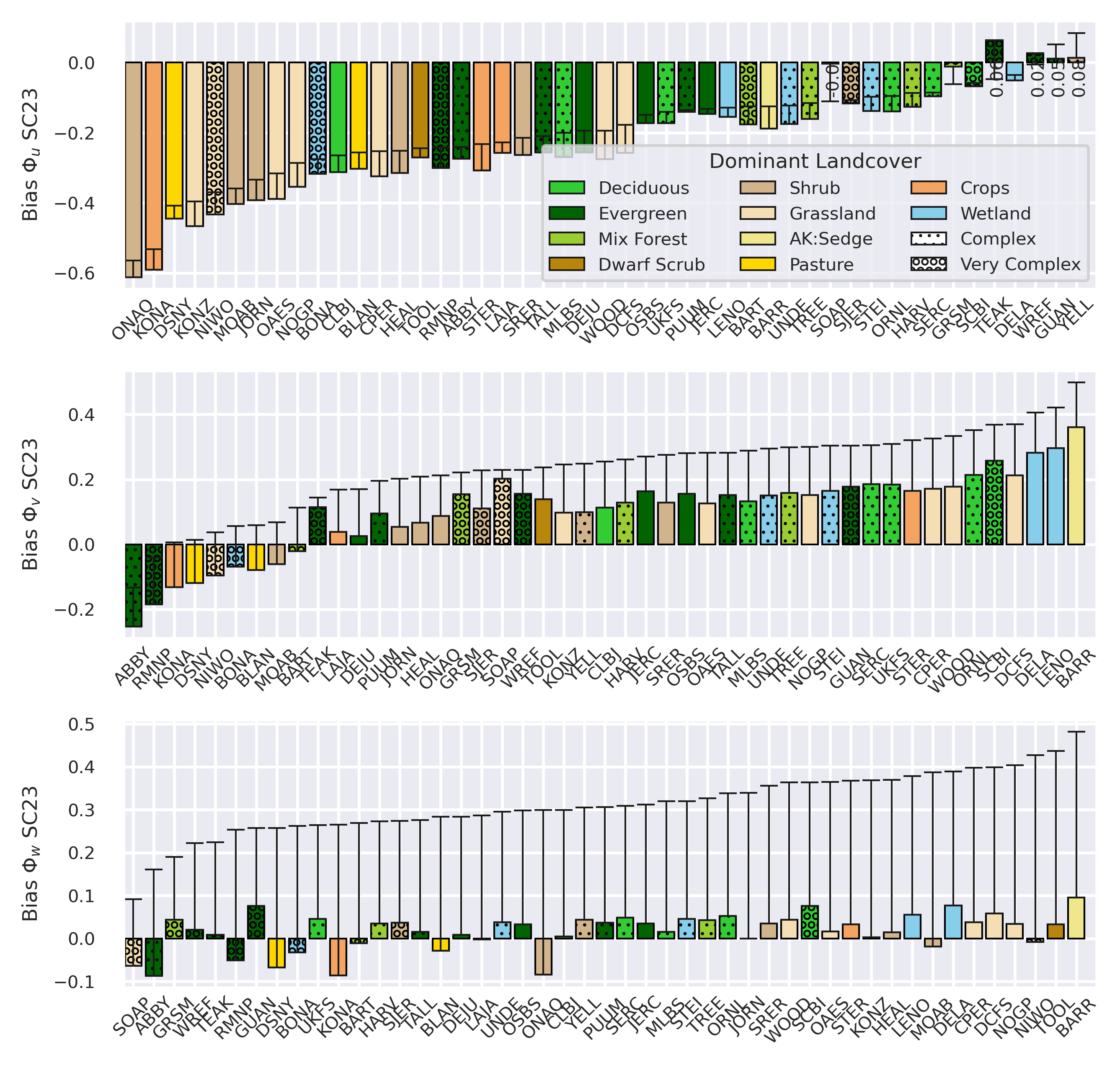}
\caption{Median bias for $\Phi_u$, $\Phi_v$ and $\Phi_w$  under stable conditions for the SC23 relations (colored bar) and for traditional MOST (black tick lines) for each site in NEON. Sites are colored according to dominant landcover and organized from lowest traditional MOST bias to highest. Sites with high complexity (median standard deviation of the digital surface model (DSM) in the flux tower footprint$>10$) are marked with black dots, and very high complexity (median standard deviation of the digital surface model (DSM) in the flux tower footprint$>20$) marked with black circles.}
\label{fig:biasvu}
\end{figure}

\begin{figure}
\centering
\includegraphics[width=5.5in]{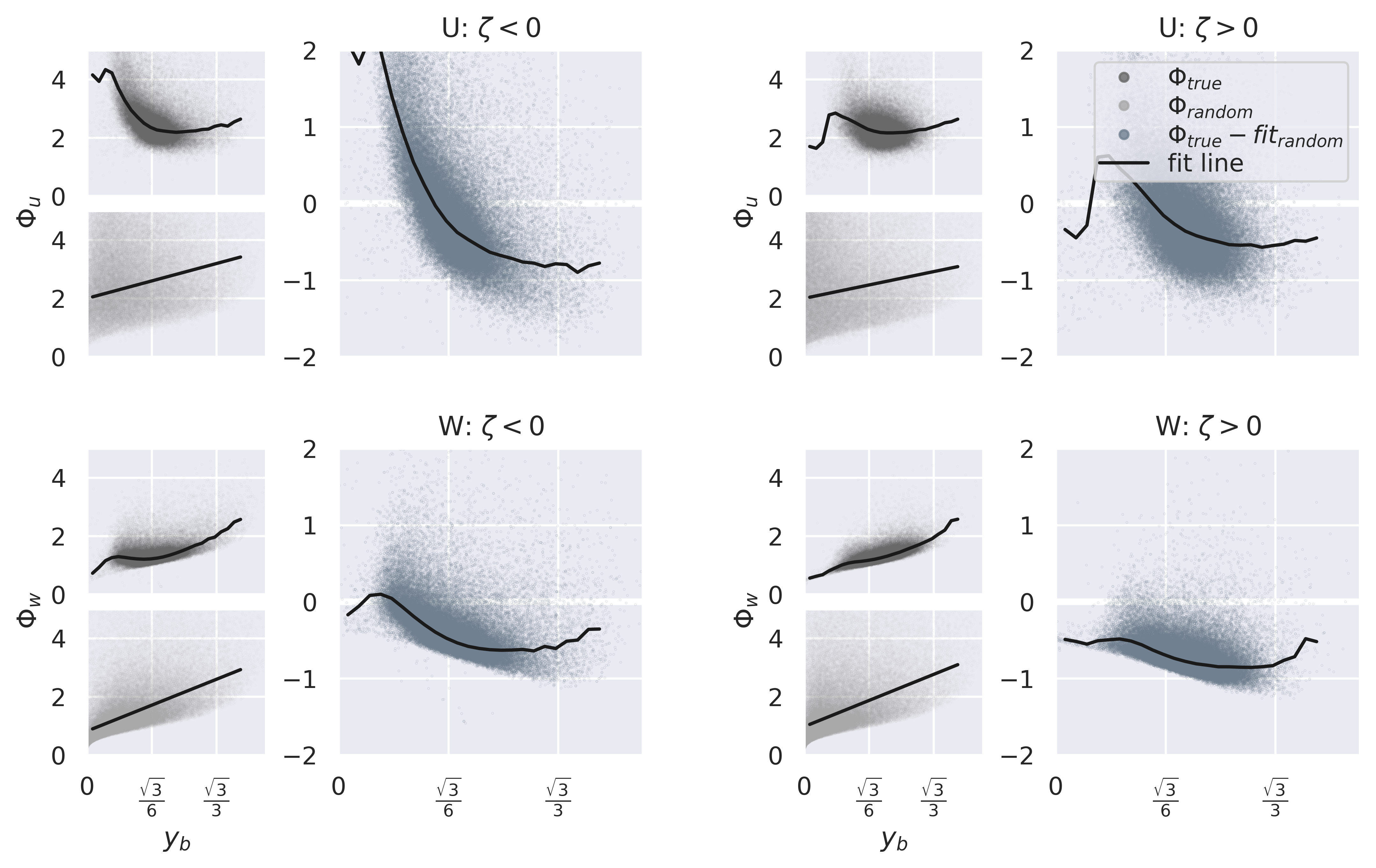}
\caption{Self similarity of $y_b$ and $\Phi_u$ under unstable conditions (a), $\Phi_u$ under stable conditions (b), $\Phi_w$ under unstable conditions (c), and $\Phi_w$ under stable conditions (d). Within each subfigure, scatterplot of the relationship between $y_b$ and $\Phi_x$ for the data with a median bin fit line (top left), the same for $y_b$ and $\Phi_x$ where the input variables ($\overline{u'u'}$, $\overline{v'v'}$, $\overline{w'w'}$, $\overline{u'v'}$, $\overline{u'w'}$, $\overline{v'w'}$) are all randomized to show degree of self similarity with a fit line in black (bottom left). Finally, scatterplot of the relationship between $y_b$ and $\Phi_x$ for the data with the fit line to the randomized data removed to show what remains once the mathematical similarity is deducted.}
\label{fig:selfsim}
\end{figure}
